\documentclass[showpacs,preprintnumbers,amsmath,amssymb,APSl,prd,nofootinbib,superscriptaddress, twocolumn]{revtex4-2}

\usepackage{dcolumn}
\usepackage{bm}
\usepackage{ifpdf}
\usepackage{hyperref}
\usepackage{bm}
\usepackage{xcolor,color,graphicx,graphics}
\usepackage[spanish,english]{babel}
\usepackage[latin1]{inputenc}
\usepackage[OT1]{fontenc}
\usepackage{latexsym,amssymb,amsmath,amsfonts}
\usepackage{makeidx}
\usepackage{epsfig,subfigure}
\usepackage{natbib}
\usepackage{epstopdf}
\usepackage{mathrsfs}
\hypersetup{colorlinks=true, linkcolor=blue, citecolor=green}
\usepackage{enumerate}
\usepackage{xcolor}
 \usepackage{multirow}
 \usepackage{ulem}

\definecolor{red}{rgb}{1,0,0}

\def\+{^\dagger}

\def\<{\leftarrow}
\def\>{\rightarrow}

\def\({\left(}
\def\){\right)}

\def\arcsinh{\mathop{\rm arcsinh}\nolimits}

\newcommand{\bi}{\begin{itemize}} 				\newcommand{\ei}{\end{itemize}}
\newcommand{\benu}{\begin{enumerate}} 		\newcommand{\enu}{\end{enumerate}}
\newcommand{\bd}{\begin{dinglist}{0}}     \newcommand{\ed}{\end{dinglist}}
\newcommand{\bfig}{\begin{figure}[htbp]}  \newcommand{\efig}{\end{figure}}
        			
\newcommand{\bc}{\begin{center}} 				  \newcommand{\ec}{\end{center}}
\newcommand{\be}{\begin{equation}} 				\newcommand{\ee}{\end{equation}}
\newcommand{\bsub}{\begin{subequations}}  \newcommand{\esub}{\end{subequations}}
\newcommand{\ben}{\begin{eqnarray}} 			\newcommand{\een}{\end{eqnarray}}
\newcommand{\ba}[1]{\begin{array}{#1}} 		\newcommand{\ea}{\end{array}}
\newcommand{\bea}{\begin{equation}\begin{array}{rcl}}
\newcommand{\eea}{\end{array}\end{equation}}

\begin{document}
\title{The optical appearance of a nonsingular de Sitter core black hole geometry under several thin disk emissions}

\author{I. De Martino} \email{ivan.demartino@usal.es}
\affiliation{Universidad de Salamanca, Departamento de Fisica Fundamental, P. de la Merced, 37008
Salamanca, Spain}

\author{R. Della Monica} \email{rdellamonica@usal.es}
\affiliation{Universidad de Salamanca, Departamento de Fisica Fundamental, P. de la Merced, 37008
Salamanca, Spain}

\author{D. Rubiera-Garcia} \email{drubiera@ucm.es}
\affiliation{Departamento de F\'isica Te\'orica and IPARCOS,
	Universidad Complutense de Madrid, E-28040 Madrid, Spain}

\date{\today}
\begin{abstract}
We consider the optical appearance under a thin accretion disk of a regular black hole with a central de Sitter core implementing $\mathcal{O}(l^2/r^2)$ far-corrections to the Schwarzschild black hole. We use the choice $l=0.25M$, which satisfies recently found constraints from the motion of the S2 star around Sgr A$^*$ in this model, and which leads to thermodynamically stable black holes. As the emission model, we suitably adapt ten samples of the Standard Unbound emission profile for a monochromatic intensity in the disk's frame, which have been previously employed in the literature within the context of reproducing General Relativistic Magneto-Hydrodynamic simulations of the accretion flow. We find the usual central brightness depression surrounded by the bright ring cast by the disk's direct emission as well as two non-negligible photon ring contributions. As compared to the usual Schwarzschild solution, the relative luminosities of the latter are significantly boosted, while the size of the former is strongly decreased. We discuss the entanglement of the background geometry and the choice of emission model in generating these black hole images, as well as the capability of these modifications of Schwarzschild solution to pass present and future tests based on their optical appearance when illuminated by an accretion disk.
\end{abstract}

\maketitle

\section{Introduction}

In the last decade direct proofs for the existence of astrophysical black holes have come from both the observations of dozens of gravitational waves out of binary black hole \cite{LIGOScientific:2016aoc} and neutron star \cite{LIGOScientific:2017vwq} mergers, and from the imaging brought by the superheated plasma in orbit around the supermassive objects at the heart of the M87 \cite{EventHorizonTelescope:2019dse} and Milky Way  \cite{EventHorizonTelescope:2022wkp} galaxies, M87* and Sgr A*, respectively. Both types of observations are compatible with the rough expectations brought by the Kerr solution of General Relativity (GR), which is solely characterized by mass and angular momentum \cite{Kerr:1963ud}. At the same time, these observations have provided a great opportunity for testing both modified black holes and black hole mimickers. The former include black holes with additional fields or hair (\textit{i.e.} hairy black holes \cite{Herdeiro:2015waa}) as well as black holes beyond GR \cite{Carballo-Rubio:2018jzw}, while the latter typically refer to horizonless compact objects of different types, such as traversable wormholes \cite{Simpson:2018tsi} or boson stars \cite{Vincent:2015xta}, see e.g. \cite{Cardoso:2019rvt} for a review of their current observational status.

A major argument in favour of these alternatives to the Kerr black hole comes from the unavoidable existence of space-time singularities inside it, as provided by the singularity theorems \cite{Senovilla:2014gza}. Such singularities are associated to the existence of at least one incomplete geodesic, and their presence threaten the classical determinism and predictability of GR itself. Since geodesic incompleteness typically correlates with the blow up of a certain set of curvature scalars, common wisdom in the field appeals to quantum-gravity regularization mechanisms when curvature nears the Planck scale and restoring, in turn, the geodesic completeness of the geometry. This way, the finding of regular black holes has become a popular trend in the community, and many different strategies to achieve this end have been concocted \cite{Lan:2023cvz}. For the sake of this paper, we consider the one of {\it de Sitter} (dS) cores, in which the central point-like singularity of a spherically symmetric (Schwarzschild) black hole is replaced by one such core, a strategy which manages to keep all curvature scalars finite \cite{Ansoldi:2008jw}. The regular de Sitter core black hole can be later set to rotate \cite{Neves:2014aba,Boshkayev:2023fft}, in order to find the counterpart of the Kerr solution.

The main aim of this work is to characterize one of these dS-core black holes via its optical appearance when illuminated by a (optically and geometrically) thin accretion disk. The chosen dS geometry is characterized by the fact that in the far-away range it introduces corrections to the Schwarzschild black hole of order $\mathcal{O}(l^2/r^2)$, where $l$ is a typical scale of the new geometry. This makes this dS core one of the strongest departures from the Schwarzschild solution within this class. Indeed, the  strong deviations in the orbits of test particles, that characterize these dS-core black holes, motivated a Monte Carlo Markov chain analysis aimed to constrain the scale length $l$ using astrometric and spectroscopic data of positions and velocity of the star S2 that orbits around the supermassive compact object Sgr A* at the centre of the Milky Way \cite{Cadoni:2022vsn, deLaurentis:2022oqa}. These data provided an upper bound on the scale length $l$ of $\lesssim 0.47M$ (in units of $G=c=1$) at the 95\% of the confidence level. For the sake of this work we shall take the choice $l=0.25M$, which lies within such a constraint and in the lower end of the branch of thermodynamically stable black holes within this class.

To carry out our analysis we shall suitably adapt ten samples of the Standard Unbound (SU) Johnson's distribution for a monochromatic (in the frame of the disk) emission profile \cite{Gralla:2020srx}, some of which have been used in several recent surveys of General Relativistic Magneto HydroDynamic (GRMHD) simulations of the accretion flow within simpler settings. Such simulations consistently report the optical appearance of black holes to be largely dominated by a bright ring of radiation enclosing a central brightness depression, the latter closely tracking, in  the observer's plane, a critical curve resulting from the projection of the unstable bound geodesics (the photon sphere), and typically known as the {\it shadow} \cite{Falcke:1999pj}. Furthermore, within this thin accretion disk framework, such a ring is broken into an infinite sequence of self-similar rings labelled by an integer $n$ corresponding to the number of half-loops of light trajectories around the black hole,  creating, on the asymptotic observer's screen, a similar number of photon rings \cite{Gralla:2019xty}. These rings produce universal signatures in very-long base interferometry (VLBI) detectors \cite{Johnson:2019ljv}, potentially allowing to distinguish between different black hole metrics \cite{Wielgus:2021peu}. However, due to the fact that such rings are exponentially decreased in their corresponding luminosities, typically those beyond $n \geq 2$ are dismissed from the images since only the $n=1$ and (perhaps) the $n=2$ ones might be observed in future observational devices, such as the next generation Event Horizon Telescope (ngEHT) \cite{Tiede:2022grp}. In addition, in this setting the outer edge of the central brightness depression is associated with the (gravitational lensed) effective region of emission (typically the horizon) and thus it can be strongly reduced as compared to the usual shadow of the geometrically thick case \cite{Chael:2021rjo}.

The features above have been confirmed in many other studies in the field, see e.g. \cite{Narayan:2019imo,Shaikh:2019hbm,Vincent:2020dij,Peng:2020wun,Gan:2021xdl,Li:2021riw,Vincent:2022fwj,Li:2022eue,Mirzaev:2022xpz,Staelens:2023jgr,Uniyal:2022vdu,Wang:2023vcv} for a non-extensive list. The dS-core black holes images studied in this work with the SU emission profiles reproduce this behaviour, but significantly decrease the luminosity extinction rate between the $n=1$ and $n=2$ photon rings, making the latter to appear comparatively much brighter in the images. At the same time, a strong reduction in the shadow's size is also observed. We discuss the interplay between the ten SU models and the dS-core background geometry in generating the photon ring and central brightness depression features of the images. We furthermore comment on the resemblances and differences of these images with respect to its Schwarzschild counterpart as well as the chances of such black holes to pass present and future observational constraints of this kind.

This work is organized as follows. In Sec \ref{Sec:II} we introduce the theoretical background and the thin accretion disk model and emission profiles used throughout the paper. Images are found and discussed in Sec. \ref{Sec:III} for both dS black holes and naked cores, and in Sec. \ref{Sec:IV} we further comment on our results. Appendix \ref{S:app} contains the main ingredients of null geodesics for the generation of images. Throughout the article we shall work in units $G=c=1$.

\section{Framework} \label{Sec:II}

\subsection{Background geometry}

We consider an asymptotically flat, spherically symmetric geometry written as
\begin{equation} \label{eq:lineel}
ds^2=-A(r)dt^2+\frac{dr^2}{A(r)}+r^2 d\Omega^2 \,,
\end{equation}
where $d\Omega^2=d \theta^2 + r^2 \sin^2 \theta d \phi^2$ is the volume element on the unit two-spheres.  A large family of dS cores parameterized by the metric function (here $l$ is a constant)
\begin{equation}
M(r)=M\left(\frac{r^q+l^q}{r^q} \right)^{\frac{p}{q}} \,,
\end{equation}
was introduced in \cite{Neves:2014aba}, since it contains as particular members the well known Bardeen \cite{Bardeen} ($p=3,q=2$) and Hayward solutions \cite{Hayward:2005gi} ($p=q=3$), and are typically found as solutions of non-linear electrodynamics (possibly coupled to scalar fields as well); the condition $p \geq 3$ must be imposed to guarantee regularity at the center $r=0$. For the sake of this work we take  the choice $p=3,q=1$, that is
\begin{equation} \label{eq:metricdS}
M(r)=M \frac{r^3}{(r+l)^3} \,.
\end{equation}
This choice represents an asymptotically flat space-time, behaving in this limit as $A(r) \approx 1- 2M/r + \mathcal{O}(l^2/r^2)$, while at the center $r \to 0$ it implements a dS core behaviour of the form $A(r) \approx 1-2Mr^2/l^3 + \mathcal{O}(r^3/l^3)$. The motivation for this model lies therefore in the fact that it provides the strongest polynomial correction to the Schwarzschild black hole at far distances within this class, while preserving the dS core at the center and the regularity of the curvature scalars it brings with it.

The metric (\ref{eq:metricdS}) represents different configurations depending on the choice of the parameter $l$. In this sense, for $0 \leq l <l_c$, where $l_c=\tfrac{8M}{27} \approx 0.2969M$, it describes two-horizons black holes with an external event horizon $r_+$ and an internal one $r_-$. These two horizons merge into a single (degenerate) one (\textit{i.e.} an extreme black hole) at $l=l_c$. For $l>l_c$ the horizon disappears, leaving a regular naked dS core. Furthermore, by analyzing the thermodynamic properties of the black holes above, one finds that the stability of the system, as given by the positivity of the specific heat (at constant $l$) can only be achieved within the range $0.245 \leq l \leq l_c$, thus leaving a narrow gap of stable black holes within this model. Furthermore, the metric \eqref{eq:metricdS} has been constrained using the astrometric and spectroscopic data of the orbital motion of the S2 star around Sgr A$^*$ \cite{Cadoni:2022vsn}.  Their analysis constrains the value of $l$ to be $ \lesssim 0.47M$ at the 95\% of the confidence level,   therefore also allowing for the existence of naked dS cores compatible with such an observation.

For our aims, the main object of interest will be the generation of images of a sample of two-horizons black hole with a dS core (dSBH), taking the choice of $l=0.25M$, which yields the event horizon radius $r_+ \approx 1.059M$ and an inner horizon radius $r_{-}=0.25M$. For completeness, we shall also discuss briefly the images of a naked dS core (NdS), for which we take the choice $l=0.30M$; obviously in such a case no horizon radius is present.

A brief description of light trajectories in spherically symmetric space-times is provided in Appendix \ref{S:app}. For our analysis below the important concept is the one of {\it photon sphere}, namely, the locus of unstable bound geodesics, and the {\it critical curve}, namely, the projection on the observer's plane of such a photon sphere.

\subsection{Accretion disk emission}

We consider our source of emission to be an optically thin, equatorial thin disk near the black hole. Such a disk is typically modelled as composed of individual emitters that follow approximately circular (Keplerian) orbits until reaching the innermost stable circular radius (ISCO) radius; after that point the particles of the disk plunge into the black hole following Cunningham's prescription \cite{Cunningham:1975zz}. We assume the disk to define a specific monochromatic intensity $I_{\nu_e}$, with $\nu_e$ the frequency in the frame of the disk, and the total intensity emitted is thus only dependent on the radial distance from the central mass, $I_{e}(r) \equiv I_{\nu_e}$. The reconstruction of the image requires, in general, solving the radiative transfer equation \cite{Gold:2020iql}
\begin{equation}
\frac{d\mathcal{I}_{\nu}}{ds} = \frac{j(\nu)}{\nu^2} -\nu \mathcal{I}_{\nu} \chi(\nu) \,,
\end{equation}
where $\mathcal{I}_{\nu} \equiv I_{\nu}/\nu^3$ is the invariant intensity, while $j_{\nu}$ is the emissivity and $\chi(\nu)$ the absorption, the latter effectively taken to zero by the optically thin assumption. The invariant character of $\mathcal{I}_{\nu}$ means that in the frame of the observer with frequency $\nu_o$ the observed intensity will behave as
\begin{equation}
I_{\nu_o}=\frac{\nu_o^3}{\nu_e^3} I_{\nu_e}= \frac{g(r_{o})}{g(r_{\infty})}  I_{\nu_e} \,,
\end{equation}
where the last equality follows from Eq.(\ref{eq:lineel}) with $g(r)=A(r)^{1/2}$. Since we are dealing with asymptotically flat space-times with an observer located far away from the black hole, then $g(r_{\infty})=1$. The total observed intensity will be then given by the integration
\begin{equation}
I_{o}(r)=\int d\nu_{o} I_{\nu_0}(r)= \int d\nu_e g^3(r)I_{\nu_e}=g^4 I(r)\,.
\end{equation}
Every time a light ray intersects the (optically thin) disk it will increase its luminosity as dictated by the disk's local intensity, and thus the final image brightness will be given by the addition of all the contributions for every possible intersection with the disk, in other words, we write
\begin{equation}
I_{o}(b)= \sum_{n} \xi_n (A^2 I)\vert_{r=r_n(b)} \,,
\end{equation}
where the function $r_n(b)$ with $n=0,1,2, \ldots$ stores the information about the radial location of the $n$-th intersection with the disk outside the event horizon, and it is dubbed as the {\it transfer function}. In this sense, $n=0$ corresponds to the {\it direct} emission of the disk, namely, those photons emitted from the disk and directly reaching the observer without undergoing any additional turns around the black hole; those with $n \geq 1$ correspond to the {\it photon rings}, i.e., photons that have completed $n$ half-turns around the black hole (since the disk is located both in the front and the back of it). Additionally, we have included a {\it fudge factor} $\xi_n$, which has been shown to increase the compatibility of the simple modelling of the emission employed here with the results of time-averaged GRMHD simulations in non-zero thickness disks \cite{Gralla:2020srx,Chael:2021rjo}; depending on the sought properties of the disk this numerical factor can be modelled according to different prescriptions. For our purposes we shall conform to the thin-disk scenario and thus take $\xi_n=1$ for $n=0,1,2$ and $\xi_n=0$ otherwise.

On theoretical grounds, the contribution of the photon rings to the total luminosity can be proven to be exponentially-suppressed. To see this, one considers a photon that starts slightly above the photon sphere $r_m$ as $r_0=r_m + \delta r_0$, where $\delta r_0 \ll r_m$. After $n$ half-turns, the photon's location will be given by \cite{Cardoso:2008bp}
\begin{equation} \label{eq:Lya}
\delta r_n = e^{\gamma_L  n} \delta r_0 \,,
\end{equation}
where $\gamma_L$ is dubbed as the {\it Lyapunov exponent}, a universal quantifier of a given background geometry in the limit $n \to \infty$. This exponential drift of the photon's location induces a similar exponential decay for the luminosities of the corresponding photon rings, hence its relevance in connecting properties of the background geometry with actual observables (up to the non-trivial modelling of the accretion disk, as we shall see below). From an observational point of view, however, a VLBI observatory can only measure luminosity contrasts between the direct emission $n=0$ and the first photon ring $n=1$, while future experiments such as the ngEHT could allow to measure up the second photon ring $n=2$ \cite{Tiede:2022grp}. Therefore, we shall keep up to the $n=2$ ring in the generation of images and the computation of the Lyapunov exponent above.

\begin{table*}[t!]
\setlength{\tabcolsep}{5pt}
\renewcommand{\arraystretch}{1.5}
\begin{tabular}{|c|c|c|c|c|}
\hline
Model & $\mu$  & $\gamma$  & $\sigma/M$  & Comments  \\ \hline
 SU1 & $2r_+$ & $-2$  & $3/2$  & Softest decay of \cite{Cardenas-Avendano:2023dzo}  \\ \hline
 SU2 & $17r_+/6$ & $-2$  & $1/4$  & This is GLM3 model of \cite{daSilva:2023jxa} for a Schwarzschild black hole.  \\ \hline
 SU3 & $\frac{3}{2} r_+$ & $2$  & $1/4$  & Narrow rings of \cite{Paugnat:2022qzy} \\ \hline
 SU4 & $r_-$ & $-2$  & $3/2$  & Overestimated of \cite{Paugnat:2022qzy}  \\ \hline
 SU5 & $r_+$ & $0$  & $1$ & Typical 1 of \cite{Paugnat:2022qzy} \\ \hline
 SU6 & $r_+$ & $2$  & $1/2$  & Underestimated of \cite{Paugnat:2022qzy} \\ \hline
 SU7 & $r_-$ & $-3/2$  & $1/2$  &  Considered in \cite{Gralla:2020srx}, GLM1 model of \cite{daSilva:2023jxa}    \\ \hline
 SU8 & $r_-$ & $0$  & $1/2$  & This is GLM2 model of \cite{daSilva:2023jxa}  \\ \hline
 SU9 & $r_-$ & $2$  & $1$  & Typical 2 of \cite{Paugnat:2022qzy} \\ \hline
 SU10 & $r_-$ & $2$ & $1/4$  & Hardest decay of \cite{Cardenas-Avendano:2023dzo}  \\ \hline
\end{tabular}
\caption{The SU class of emission profiles defined in Eq.(\ref{eq:SU}) for the ten choices of the parameters $\mu$, $\gamma$, and $\sigma/M$ characterizing them. Here $r_{\pm}$ denote the horizon and inner radius of the black hole; for Schwarzschild this is $r_+=2M$ and $r_{-}=0$, while for dSBH this is $r_+ \approx 1.059M$ and $r_-=0.25M$. In the rightmost column we display the references each model is inspired from.}
\label{Table:I}
\end{table*}

\begin{figure*}[t]
\includegraphics[width=8.2cm,height=5.5cm]{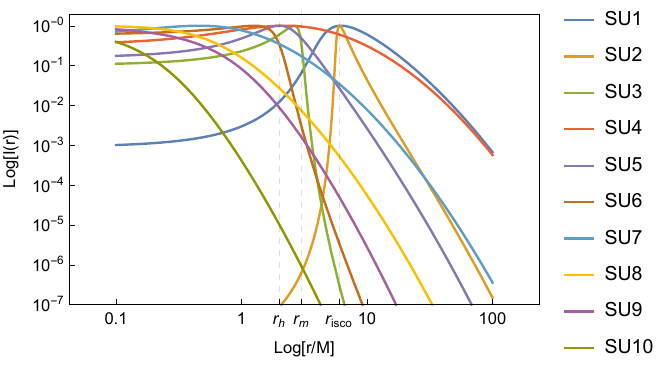}
\includegraphics[width=8.2cm,height=5.5cm]{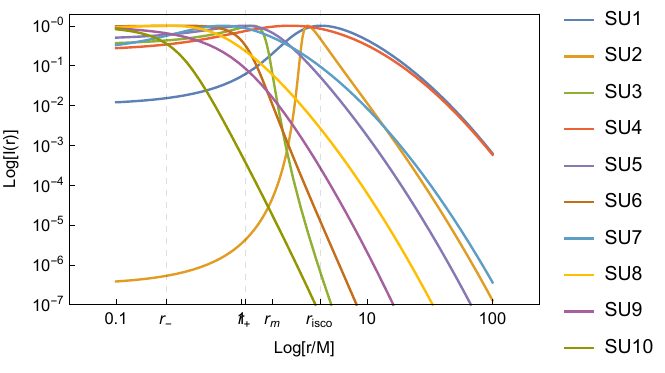}
\caption{The ten (normalized) emission profiles of Table \ref{Table:I} for the Schwarzschild (left) and dSBH (right) configurations in double-logarithmic scale. The three dashed vertical lines denote the event horizon radius $r_+$ ($r_h$ in the Schwarzschild case representing the single horizon in this geometry), the photon sphere radius $r_m$  and the ISCO radius $r_{\rm ISCO}$, respectively, for each configuration; for the dSBH a fourth vertical dashed line denotes the location of the inner horizon radius $r_-$. Obviously, only those parts of the emission profile at radius $r>r_+$ are relevant for the generation of images.}
\label{fig:emission}
\end{figure*}

As our class of emission profiles we select Johnson's SU models, given by the expression \cite{Gralla:2020srx}
\begin{equation} \label{eq:SU}
J_{SU}=\frac{e^{-\frac{1}{2} \left[\gamma +\arcsinh \left(\frac{r-\mu}{\sigma}\right) \right]^2}}{\sqrt{(r-\mu)^2+\sigma^2}}\,,
\end{equation}
which is characterized by three parameters; $\mu$ is related to the location of the peak of the emission: small (big) values tend to move the peak away from (behind) the event horizon; $\gamma$ governs the profile asymmetry: negative (positive) values tend to locate the steep part of the profile inside (outside) the horizon; $\sigma$ (typically given in units of $M$) is related to the profile width; small (big) values tend to have a steeper (broader) decay with narrower (broader) rings. For the sake of this work we shall take ten choices for these values covering a range of qualitatively different profiles, and which are taken from a number of different sources where the parameters range from $\mu \in [r_-,2r_+]$, $\gamma \in [-2,2]$ and $\sigma/M \in [0.25,1.5]$, where $r_{\pm}$ are the outer and inner horizons of a rotating black hole. Despite the lack of rotation in our model, we shall use the outer/inner horizons of the dSBH configuration as a substitute for them, while in the Schwarzschild black hole where only the event horizon is present, we obviously have $r_+=2M$ and $r_-=0$.

\begin{figure*}[t!]
\includegraphics[width=5.4cm,height=4.6cm]{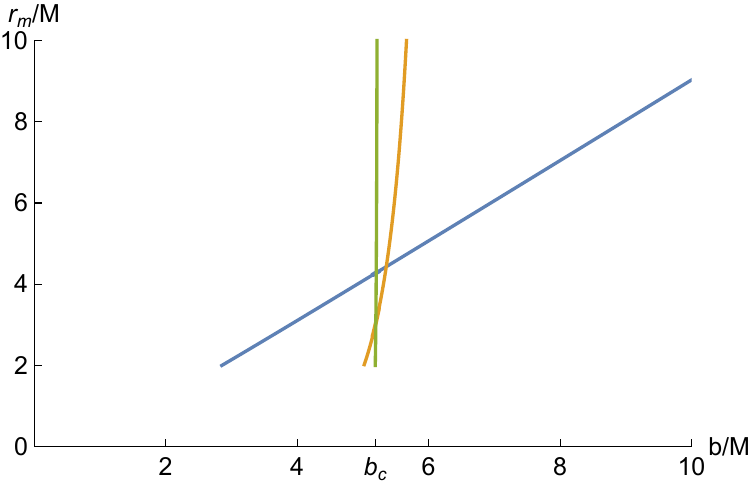}
\includegraphics[width=5.4cm,height=4.6cm]{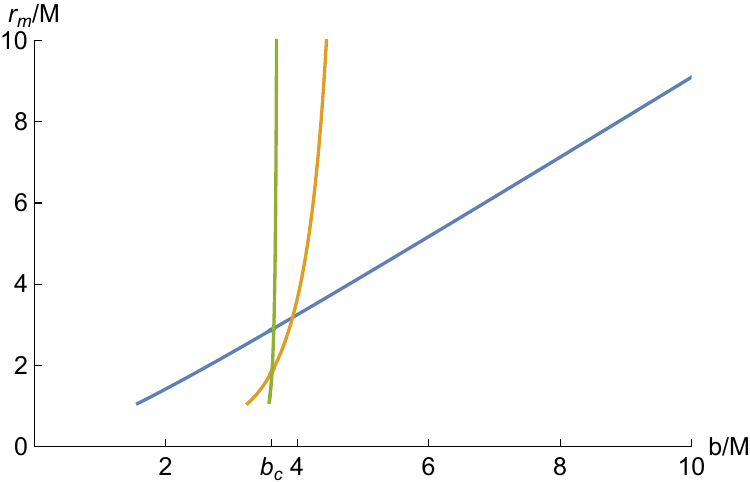}
\includegraphics[width=5.4cm,height=4.6cm]{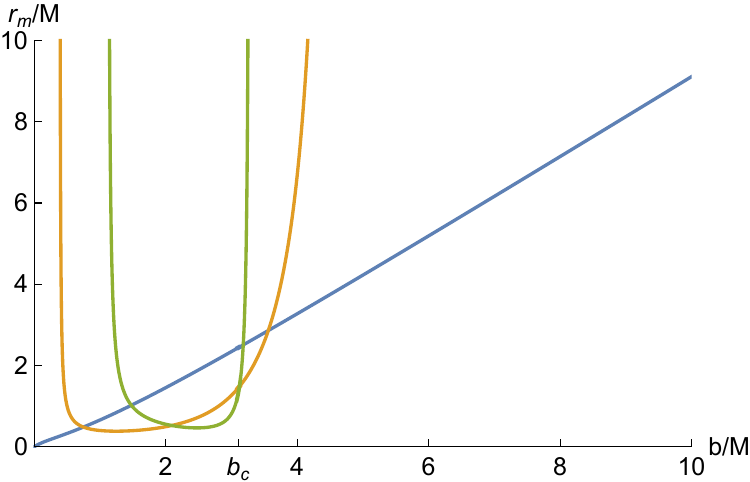}
\caption{The transfer function $r_n(b)$ for the Schwarzschild black hole (left), dSBH (middle) and NdS (right) solutions. The contributions of the direct emission correspond to blue straight lines, while the $n=1$ and $n=2$ photon rings are shown in orange and green, respectively.}
\label{fig:transfunc}
\end{figure*}

This pool of parameters is detailed in Table \ref{Table:I}, where the different models are presented in decreasing order in terms of the radius $r_e$ of the region where they possess the effective source of emission \textit{i.e.} its maximum. The corresponding radial dependence of these profiles is depicted in Fig. \ref{fig:emission} for the Schwarzschild black hole (left) and the dSBH (right), including their respective values of the event horizon radius $r_+$, the inner one $r_-$ (in the dSBH case), the photon sphere radius $r_m$ (associated to the locus of unstable geodesics, see Appendix \ref{S:app}) and the ISCO. Three of these choices are adapted from those originally introduced by Gralla, Marrone and Lupsasca (GLM) in Ref. \cite{Gralla:2020srx}, and employed  for the generation of images in several papers by some of us, such as in \cite{daSilva:2023jxa}; here they are dubbed as SU2, SU7, and SU8, respectively. Five of the remaining models were introduced by Paugnat et al. in \cite{Paugnat:2022qzy} as representative cases of a survey performed over black hole spins and inclinations for combination of values of the emission profile parameters $\mu=\{r_-,r_+/2,r_+,3r_+/2,2r_+\}$, $\gamma=\{-2,-1,0,1,2\}$ and $\sigma/M=\{0.25,0.5,1,1.5\}$,  which cover the range of results found in the corresponding simulations performed by such a collaboration. Such models represent a ``Typical 1" profile with a $n=2$ ring diameters slightly above average values (SU5), a ``Typical 2" profile with diameters slightly below average values (SU9), an overestimated profile (SU4) and an underestimated one (SU6), and finally a narrow profile with the smallest diameters (SU3). The remaining two choices come from the analysis of \cite{Cardenas-Avendano:2023dzo} and, in particular, from their Table 2, in which they present over a hundred of emission profiles, highlighting three of them. The first one (SU1) corresponds to the largest radius of emission and a slow decay, while the second one (SU10) lies in the opposite end, with a very quick decay. The third model is the GLM1 one already included in our list as the SU7 one.

This pool of models is flexible enough to cover different scenarios of behaviour of the radial dependence around the effective region of emission of the disk. In this sense, four models peak outside the event horizon of the Schwarzschild black hole: SU1 and SU2 near the ISCO radius with soft and hard decays, respectively, and SU3 and SU4 slightly above the Schwarzschild radius, with hard and soft decays, respectively. Another one (SU5) peaks almost exactly at $r=r_+$. The remaining SU6/SU7/SU8/SU9/SU10 models peak inside the event horizon, so the effective radius of emission is marked by their values at $r=r_+$. These last five models have increasingly harder decays, which means that for those in the end of the sequence (SU8, SU9, and SU10), there will be little emission outside the event horizon, something that should have its reflection in the corresponding optical appearances. Given the fact that $r_{\pm}$ are different in the Schwarzschild/dSBH solutions, there are some differences in the shape of the corresponding profiles; this seems to be an unavoidable trouble of comparing different background geometries: the behaviour of the disk is {\it not} independent of the shape of the metric, see \cite{Boshkayev:2023fft} for a recent discussion on this. Accepting this, we normalize all our intensity profiles to their maximum values for as an optimal comparison of the Schwarzschild/dSBH images as possible.

\section{Generation of images} \label{Sec:III}

\begin{table*}[t!]
	\setlength{\tabcolsep}{3pt}
	\renewcommand{\arraystretch}{3}
	\begin{tabular}{|c|c|c|c|c|c|c|c|c|c|c|c|c|}
		\hline
		Model & $\gamma_L$ & $e^{\gamma_L}$   & $\left(\frac{I_1}{I_2}\right)^{SU1}$ &  $\left(\frac{I_1}{I_2}\right)^{SU2}$ & $\left(\frac{I_1}{I_2}\right)^{SU3}$ & $\left(\frac{I_1}{I_2}\right)^{SU4}$ & $\left(\frac{I_1}{I_2}\right)^{SU5}$ & $\left(\frac{I_1}{I_2}\right)^{SU6}$ & $\left(\frac{I_1}{I_2}\right)^{SU7}$ & $\left(\frac{I_1}{I_2}\right)^{SU8}$ & $\left(\frac{I_1}{I_2}\right)^{SU9}$ & $\left(\frac{I_1}{I_2}\right)^{SU10}$ \\ \hline \hline
		Sch  & $3.1507$  & $23.35$  & $28.94$  & $27.83$ & $21.57$  & $27.21$ & $23.34$  & $21.14$  & $24.74$  & $23.45$  & $22.81$  & $22.20$  \\ \hline
		dSBH   & $2.266$  & $9.64$  & $14.06$  & $12.51$  & $8.00$  & $13.14$  & $10.07$  & $8.20$  & $10.82$ & $9.61$  & $9.14$  & $8.47$ \\ \hline
	\end{tabular}
	\caption{The Lyapunov index $\gamma_L$, the theoretical extinction rate $e^{\gamma_L}$, and the observational extinction rate $I_1/I_2$ between the $n=1$ and $n=2$ photon rings for the ten SU models listed in Table \ref{Table:I}, as obtained for the Schwarzschild and the dSBH models.}
\label{Table:II}
\end{table*}

The observational appearances of the Schwarzschild, dSBH and NdS configurations  are built using a ray-tracing procedure: all trajectories arriving to the observer's plane image are backtracked into the emitting region via the geodesic equation (recall that a short derivation of the equations of null geodesics is provided in the Appendix \ref{S:app})
\begin{equation} \label{eq:dphidr}
\frac{d\phi}{dr}=-\frac{b}{r^2} \frac{1}{\sqrt{1-b^2 \frac{A(r)}{r^2}}}\,.
\end{equation}
The relevant quantities to carry out this analysis and to characterize the resulting features are the critical impact parameter and photon sphere radius. For Schwarzschild these are $b_c=3\sqrt{3}M \approx 5.196M$ and $r_m=3M$, while for dSBH $b_c \approx 3.614M$ and $r_m \approx 1.767M$, and, finally, for NdS one has $b_c \approx 3.108M$ and $r_m \approx 1.329M$.

We place the observer's at a radius of $r_{\infty}=1000M$ to recover the asymptotically Schwarzschild character of the space-time there,  and run the integrations to collect the info on the transfer function, which is depicted in Fig. \ref{fig:transfunc} for the Schwarzschild (left), dSBH (middle) and NdS (right) configurations; to this end we assume the disk to be seen in a face-on orientation. This figure displays the contributions of the direct emission (blue straight line) and the $n=1$ (orange) and $n=2$ (green) photon rings: the effective source of emission $r_e/M$ determines the parts of these curves above which their luminosity can actually reach the asymptotic observer. Since the slopes of the curves are associated with the demagnification of their corresponding emissions in the images, this figure states the dominant role of the direct emission in the image via its nearly constant slope; the $n=1$ and $n=2$ photon rings will be sub-dominant, while $n>2$ rings will be ignored. In the NdS case we observe that both photon rings extend much deeper into the inner region of the configuration; this is due to the absence of an event horizon which allows photons to reach the innermost regions of the effective potential, where an infinite divergence is found, thus repelling light rays that get to it. We shall see in a while the repercussions of this fact for the corresponding optical appearances.

\subsection{Regular dS black hole vs Schwarzschild}

\begin{figure*}[t!]
\includegraphics[width=4.4cm,height=4cm]{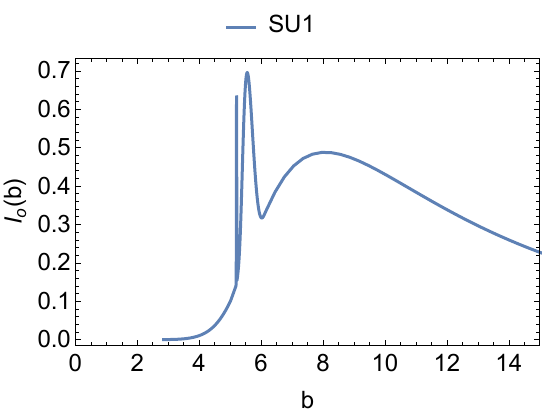}
\includegraphics[width=4.4cm,height=4cm]{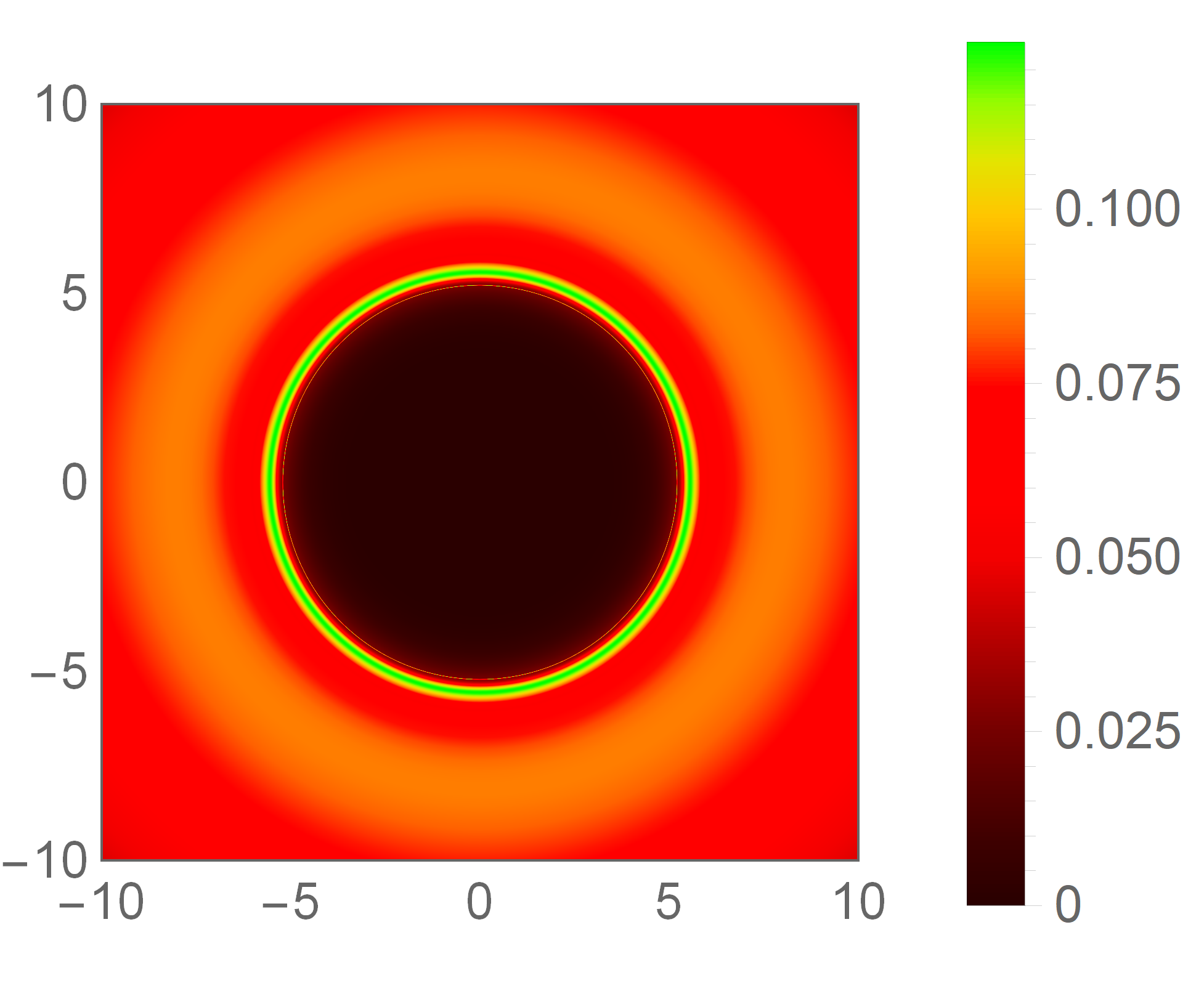}
\includegraphics[width=4.4cm,height=4cm]{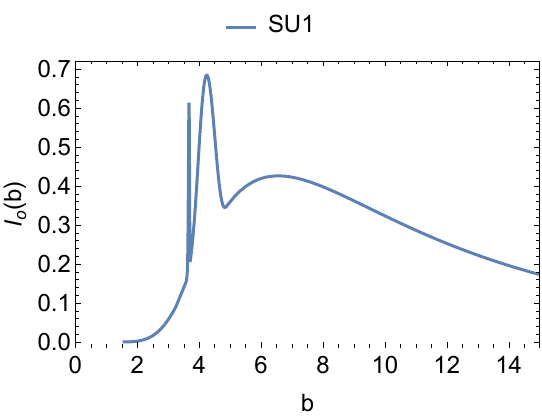}
\includegraphics[width=4.4cm,height=4cm]{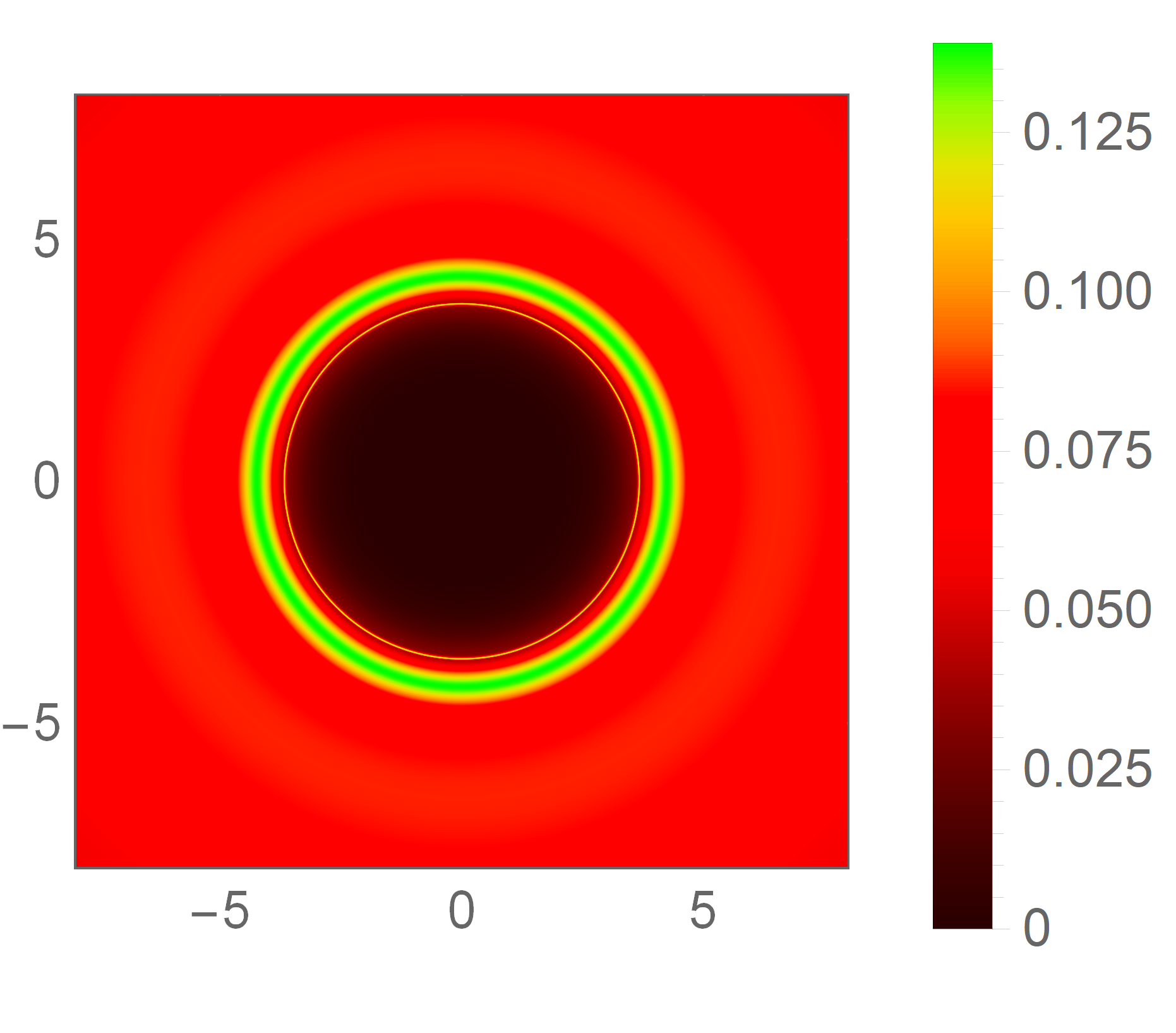} \\
\includegraphics[width=4.4cm,height=4cm]{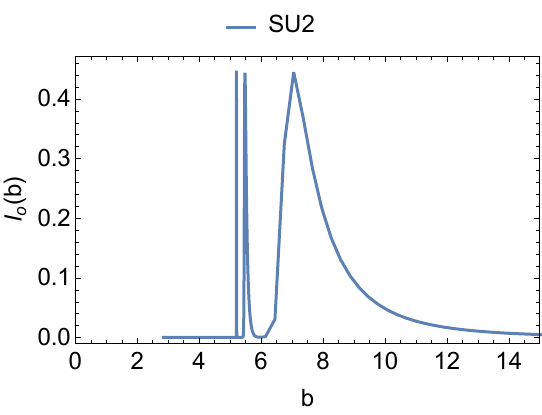}
\includegraphics[width=4.4cm,height=4cm]{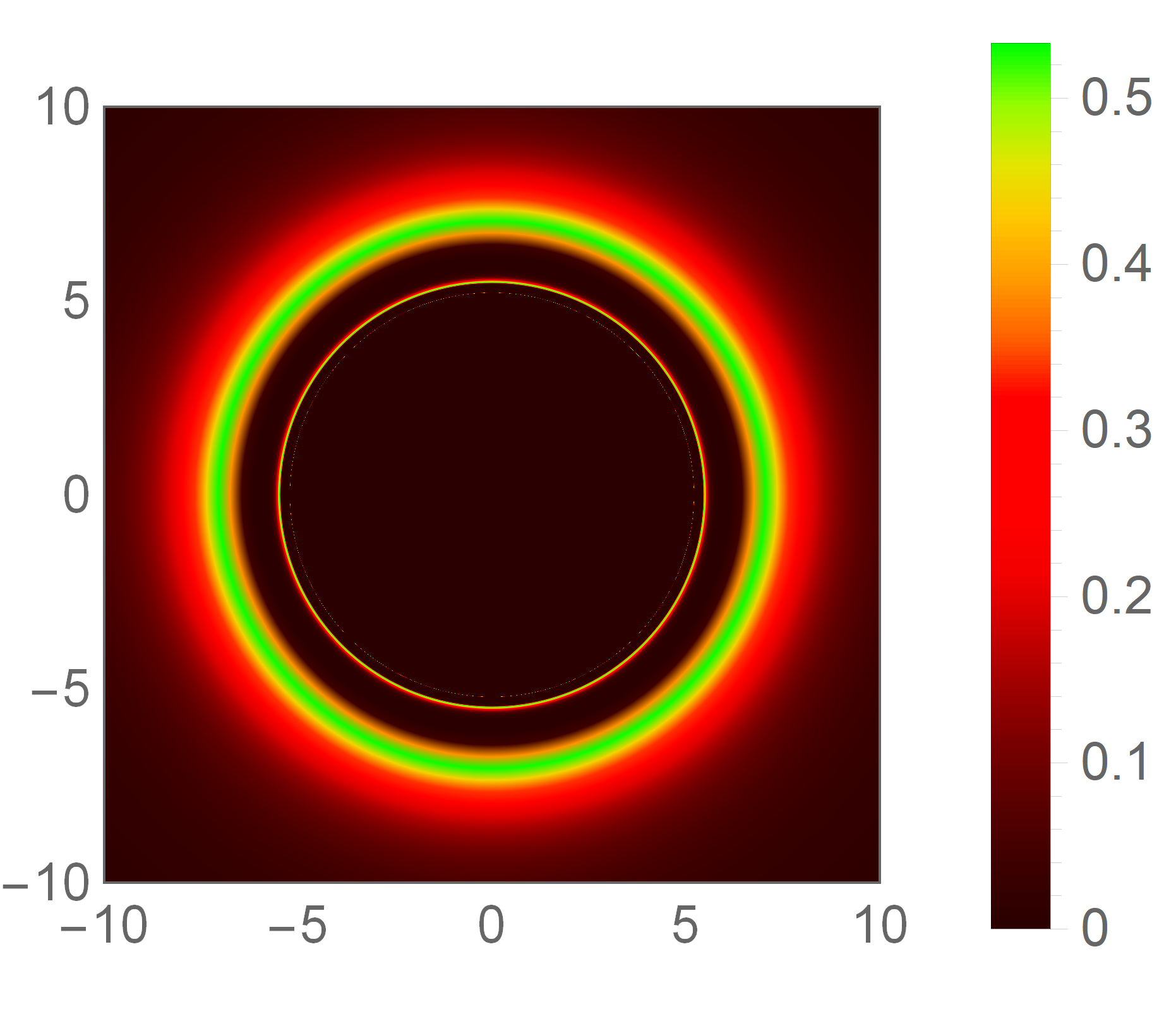}
\includegraphics[width=4.4cm,height=4cm]{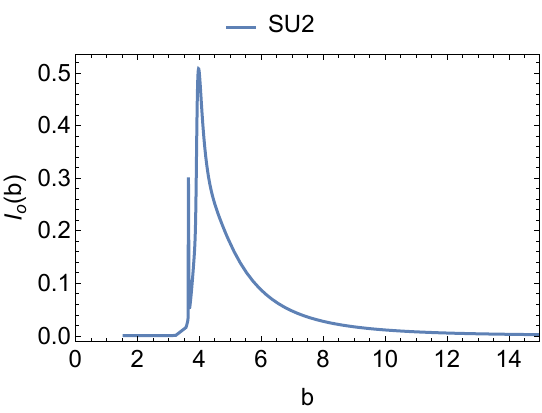}
\includegraphics[width=4.4cm,height=4cm]{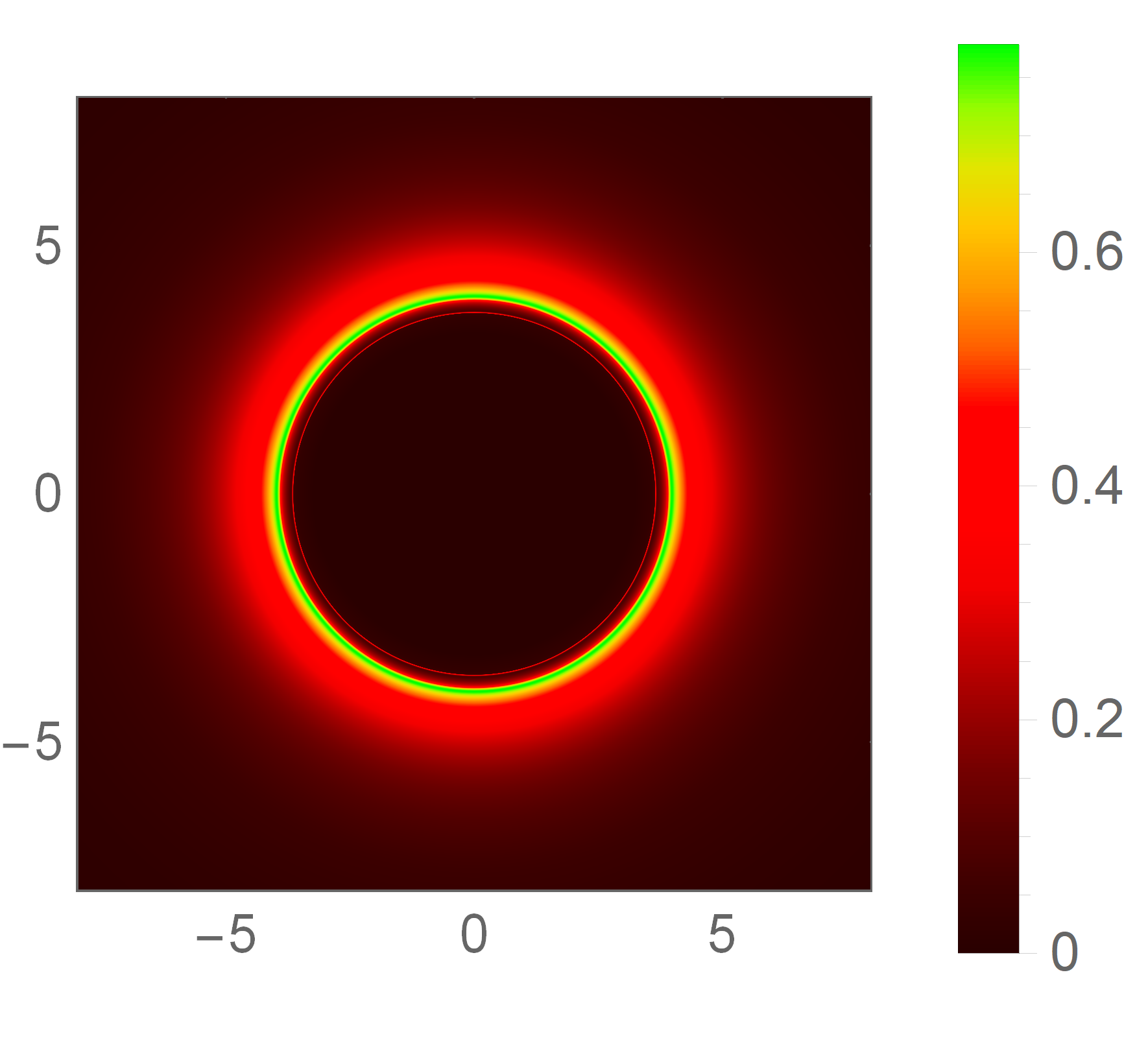} \\
\includegraphics[width=4.4cm,height=4cm]{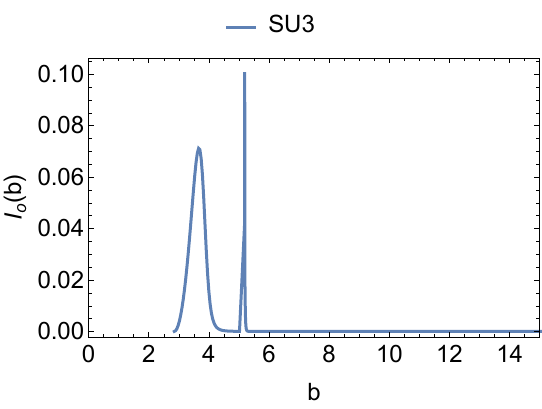}
\includegraphics[width=4.4cm,height=4cm]{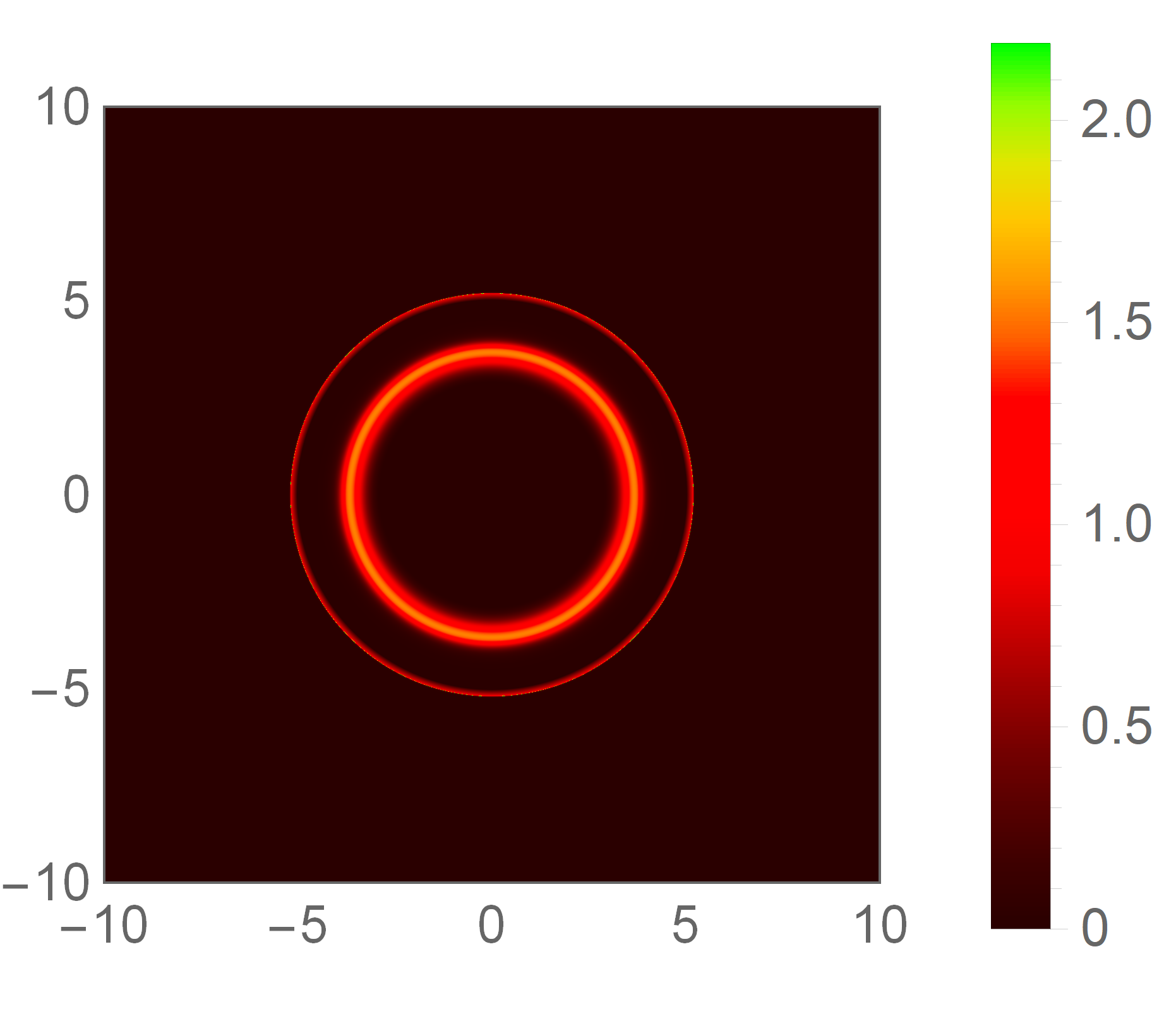}
\includegraphics[width=4.4cm,height=4cm]{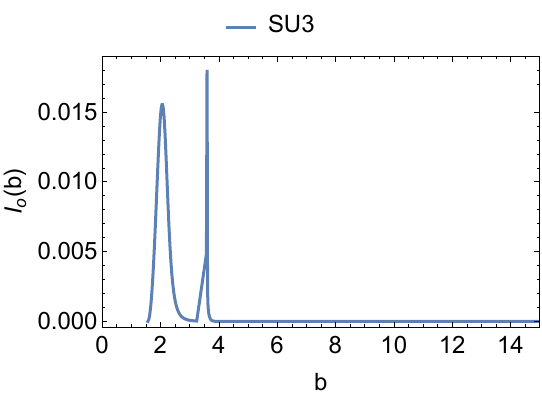}
\includegraphics[width=4.4cm,height=4cm]{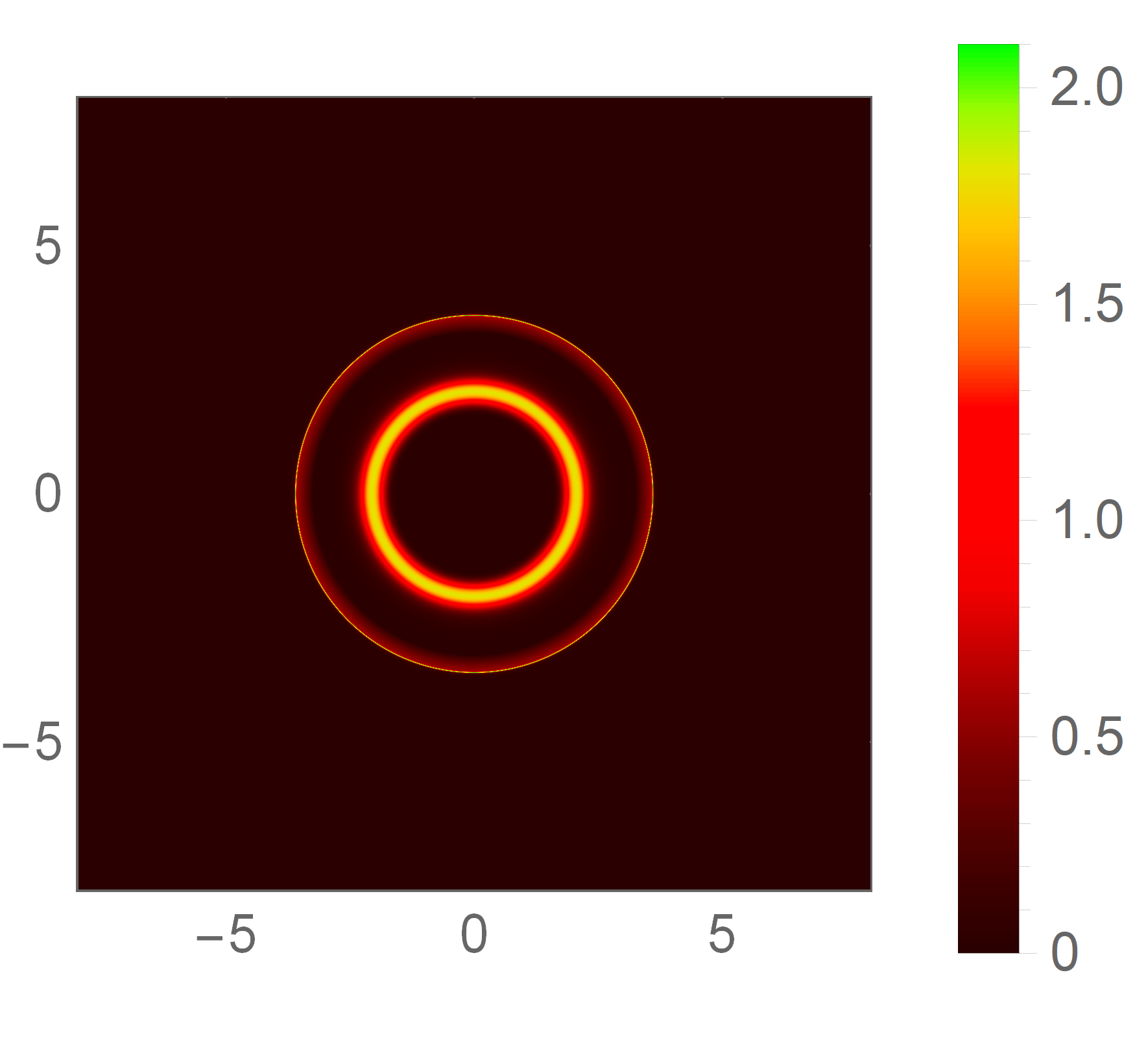} \\
\includegraphics[width=4.4cm,height=4cm]{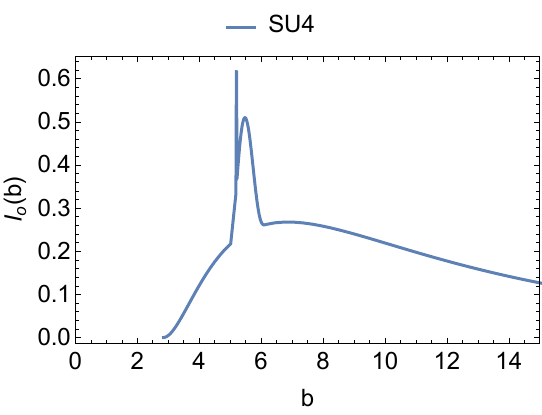}
\includegraphics[width=4.4cm,height=4cm]{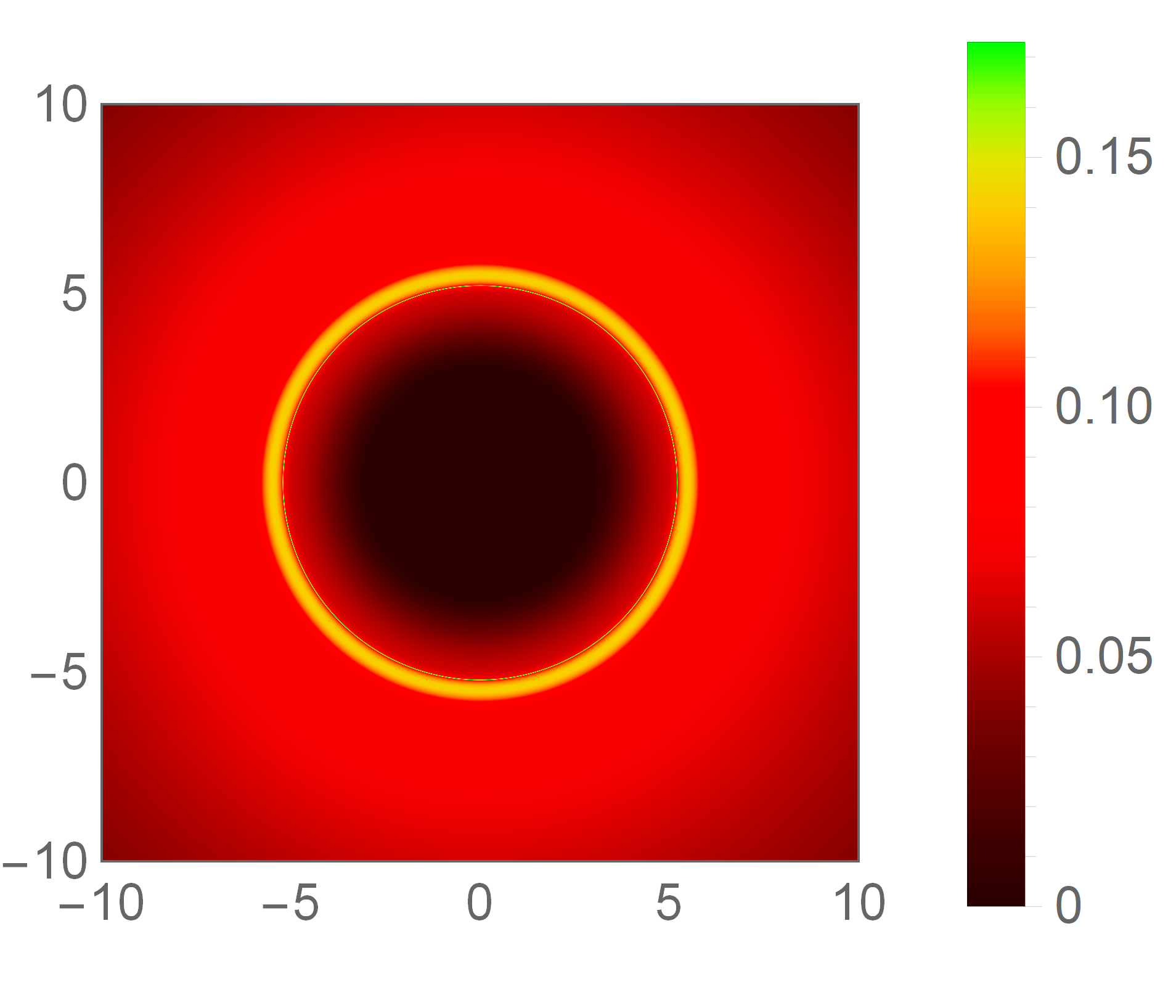}
\includegraphics[width=4.4cm,height=4cm]{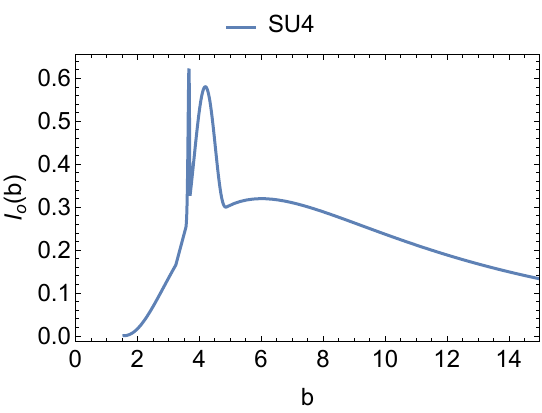}
\includegraphics[width=4.4cm,height=4cm]{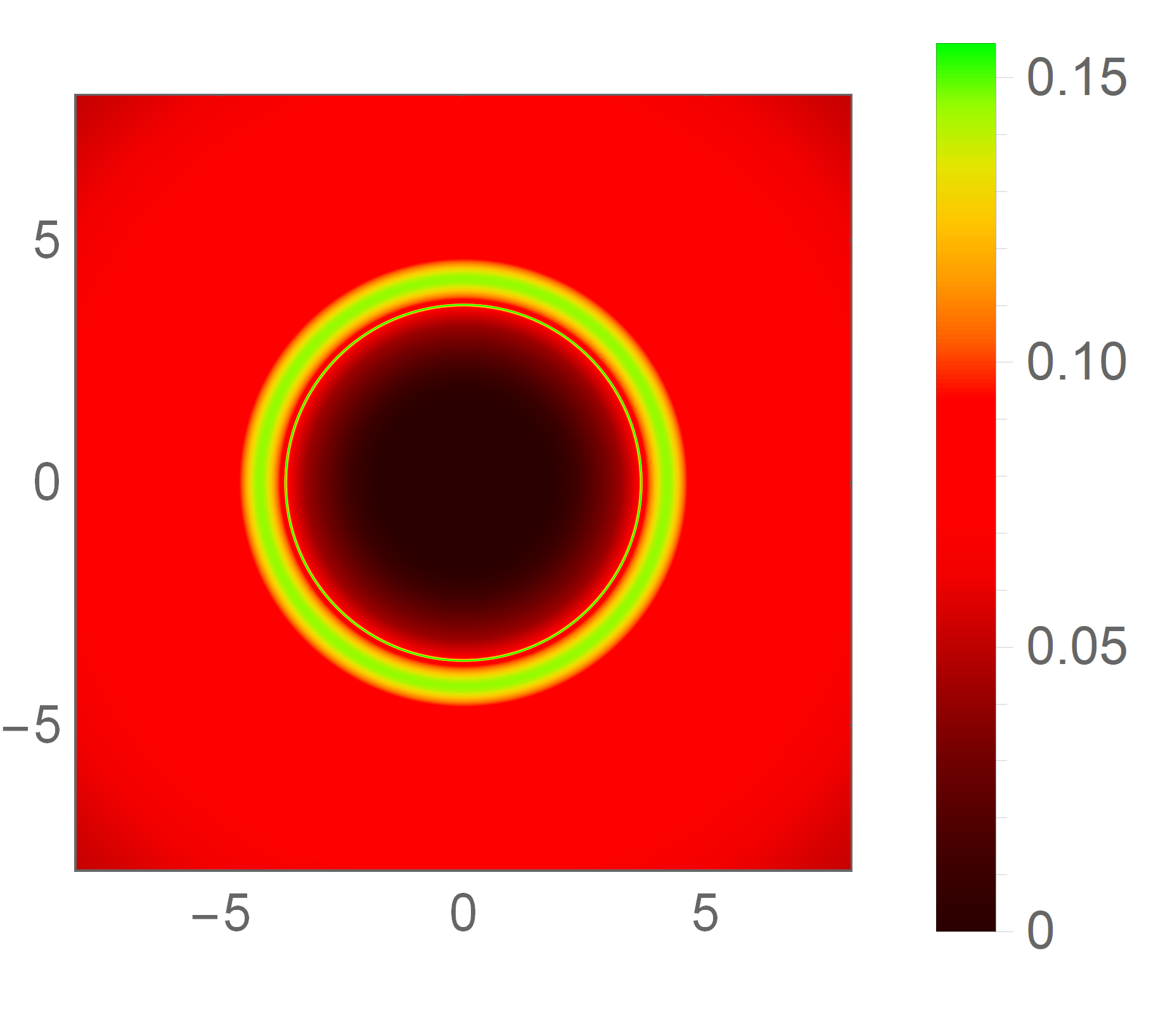} \\
\includegraphics[width=4.4cm,height=4cm]{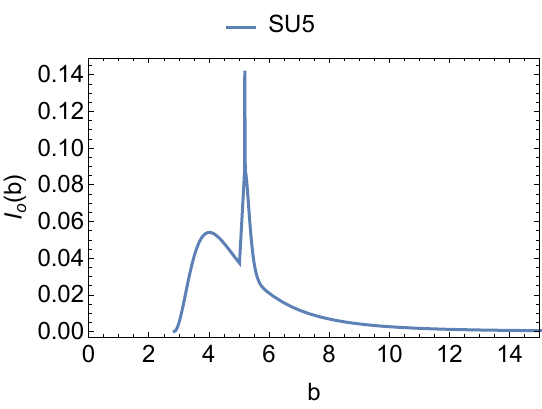}
\includegraphics[width=4.4cm,height=4cm]{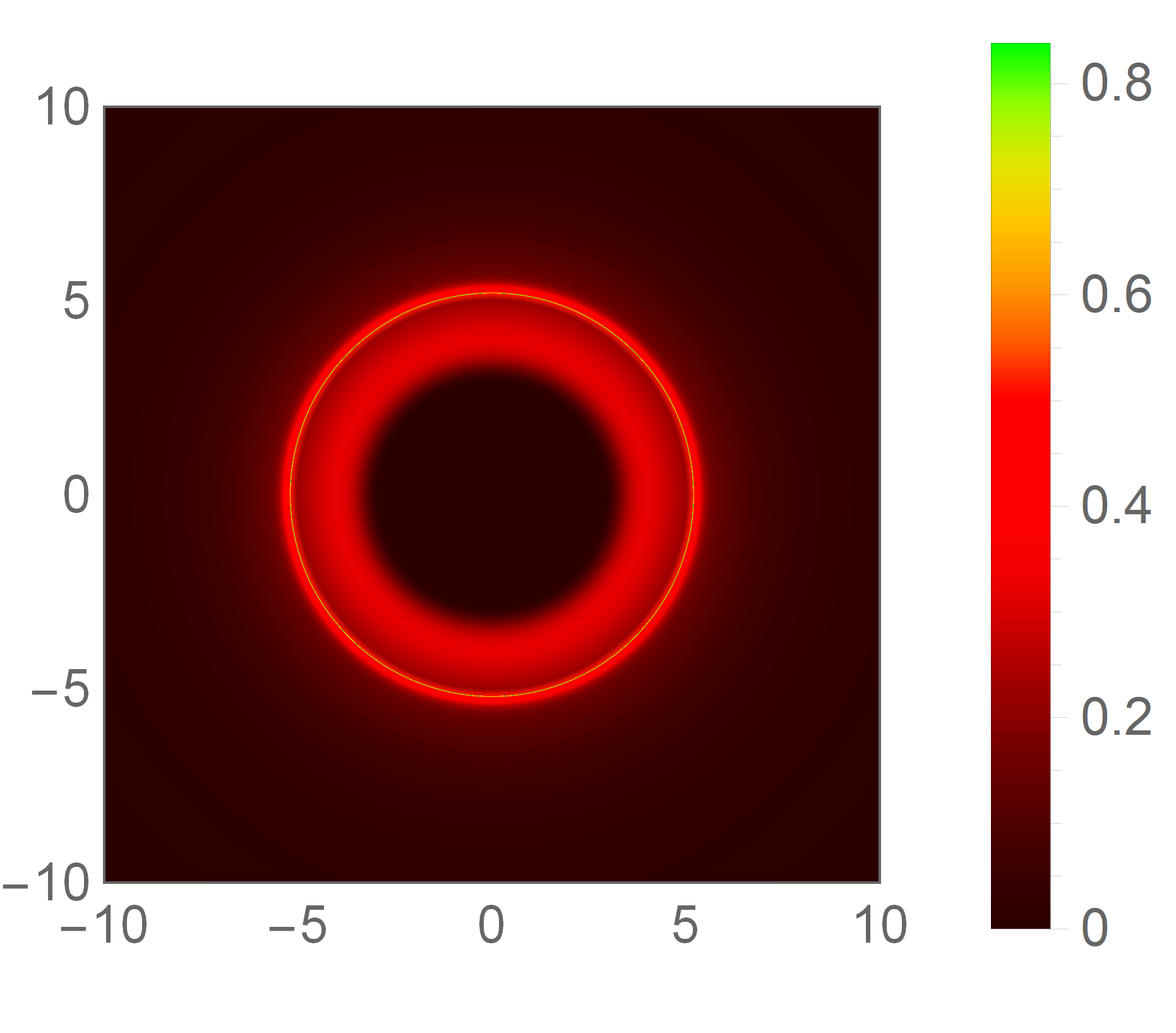}
\includegraphics[width=4.4cm,height=4cm]{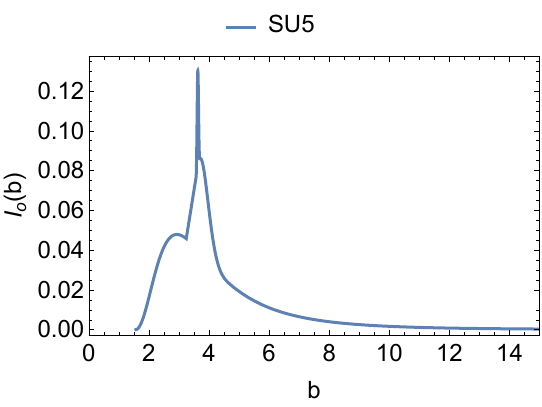}
\includegraphics[width=4.4cm,height=4cm]{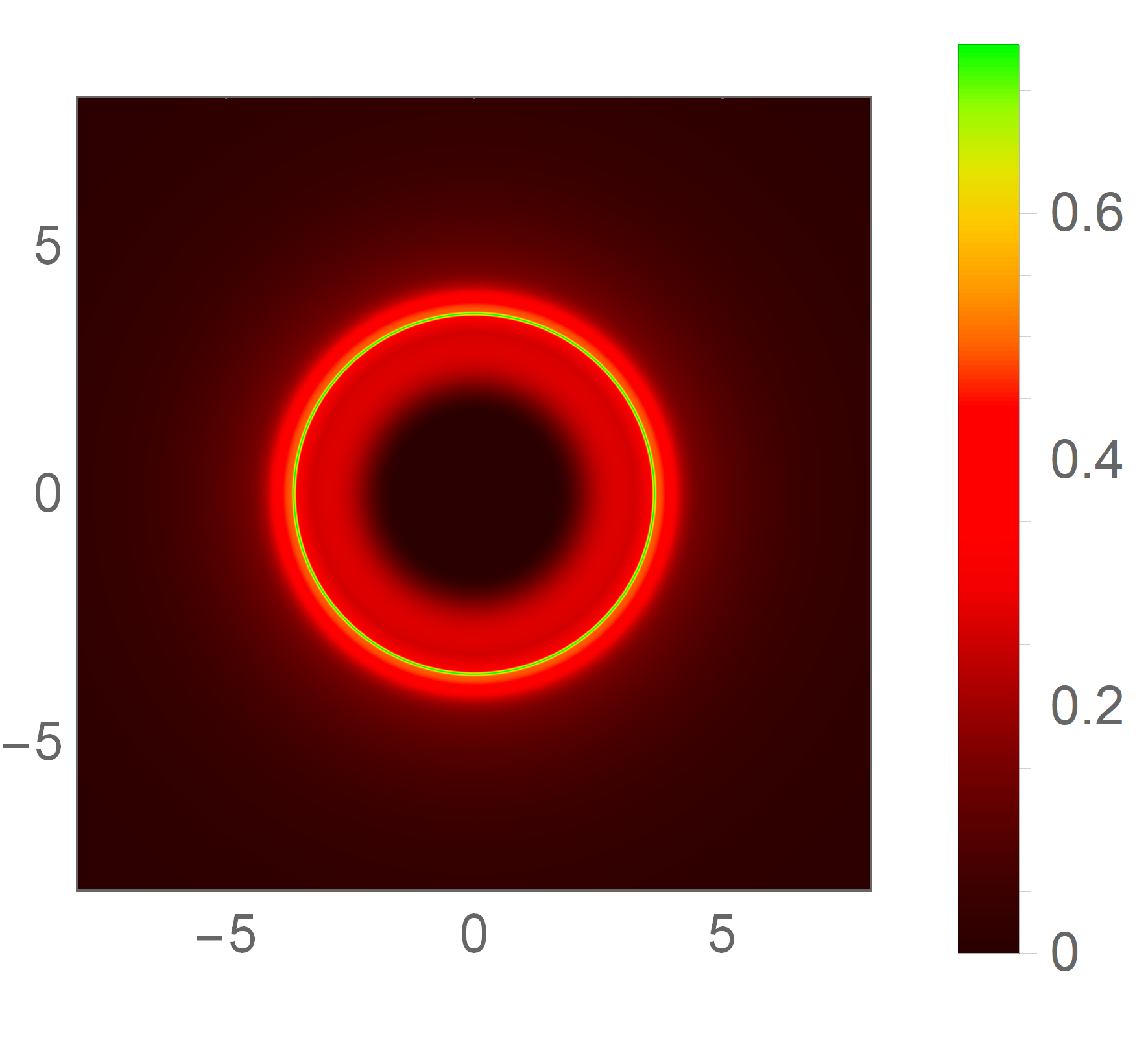}
\caption{The observed intensity $I_{o}(b)$ and the optical appearance (using $b \in [-8,8]$) of a Schwarzschild black hole (left two figures) and a dSBH solution (right two figures) for the (from top to bottom) SU1, SU2, SU3, SU4, and SU5 emission models of Table \ref{Table:I}, using a fudge factor $\xi_0=\xi_1=\xi_2=1$ and $\xi_n=0$ for $n>2$.}
\label{fig:images1}
\end{figure*}

\begin{figure*}[t!]
\includegraphics[width=4.4cm,height=4cm]{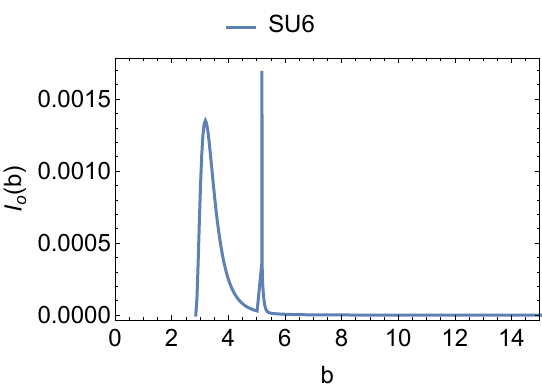}
\includegraphics[width=4.4cm,height=4cm]{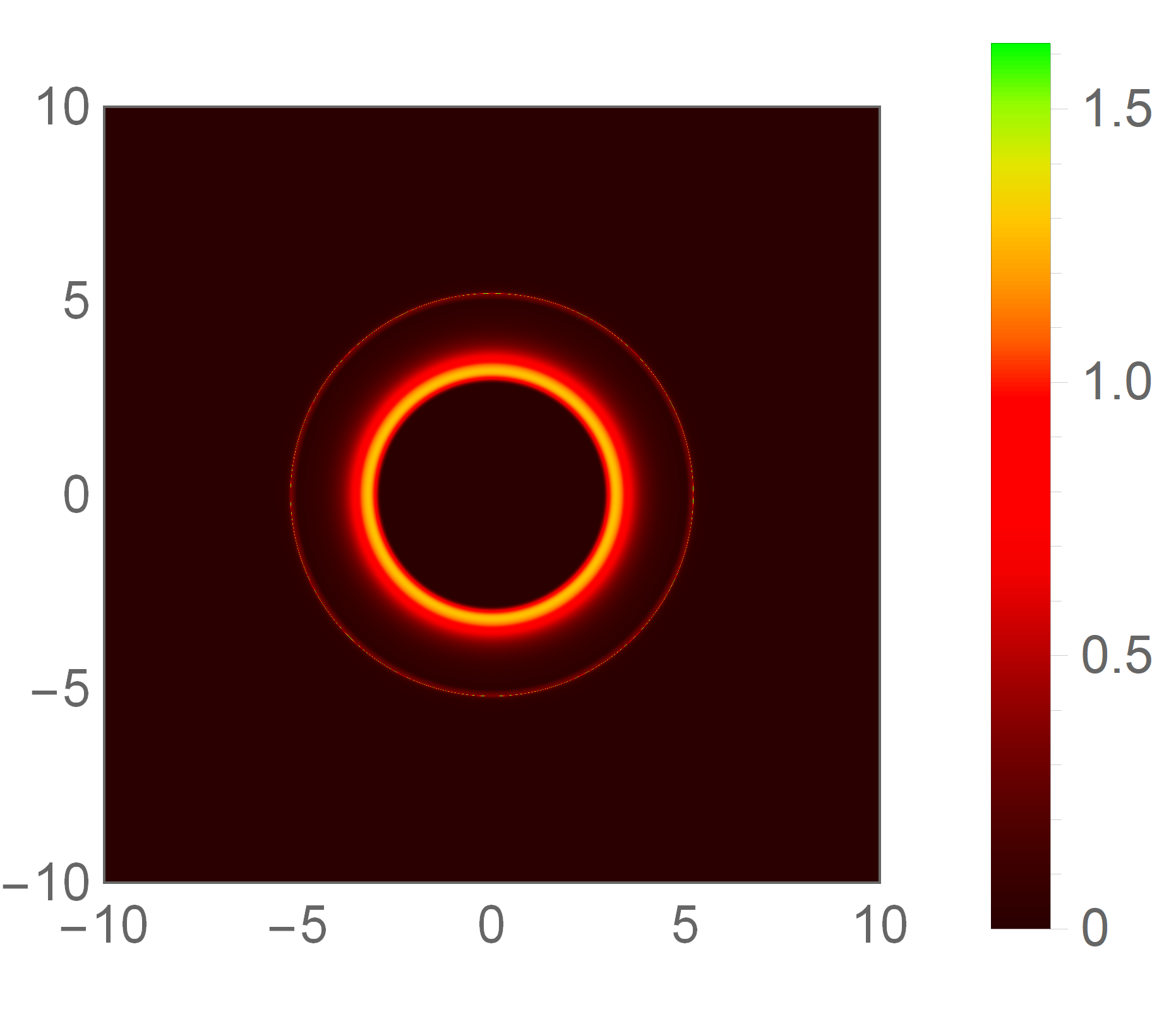}
\includegraphics[width=4.4cm,height=4cm]{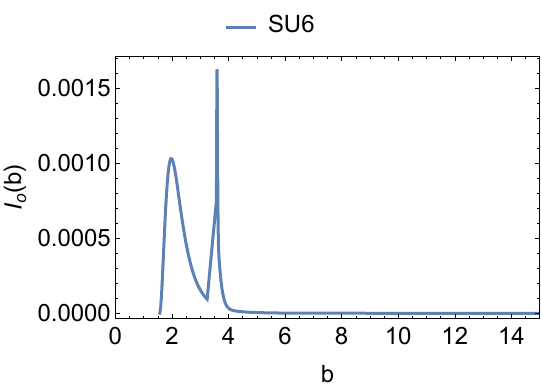}
\includegraphics[width=4.4cm,height=4cm]{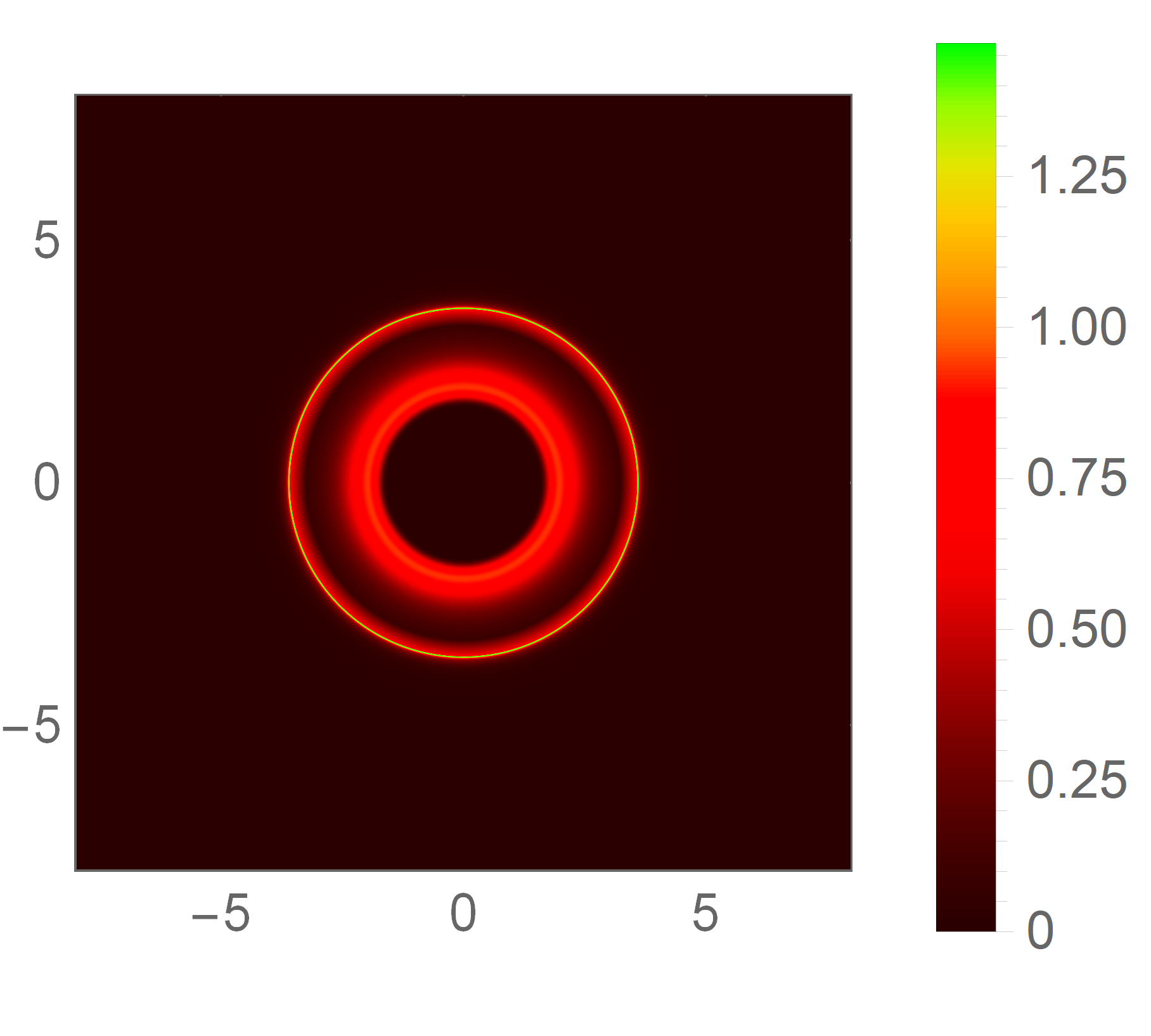} \\
\includegraphics[width=4.4cm,height=4cm]{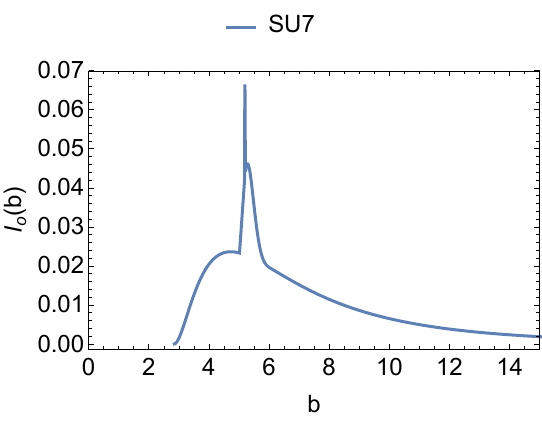}
\includegraphics[width=4.4cm,height=4cm]{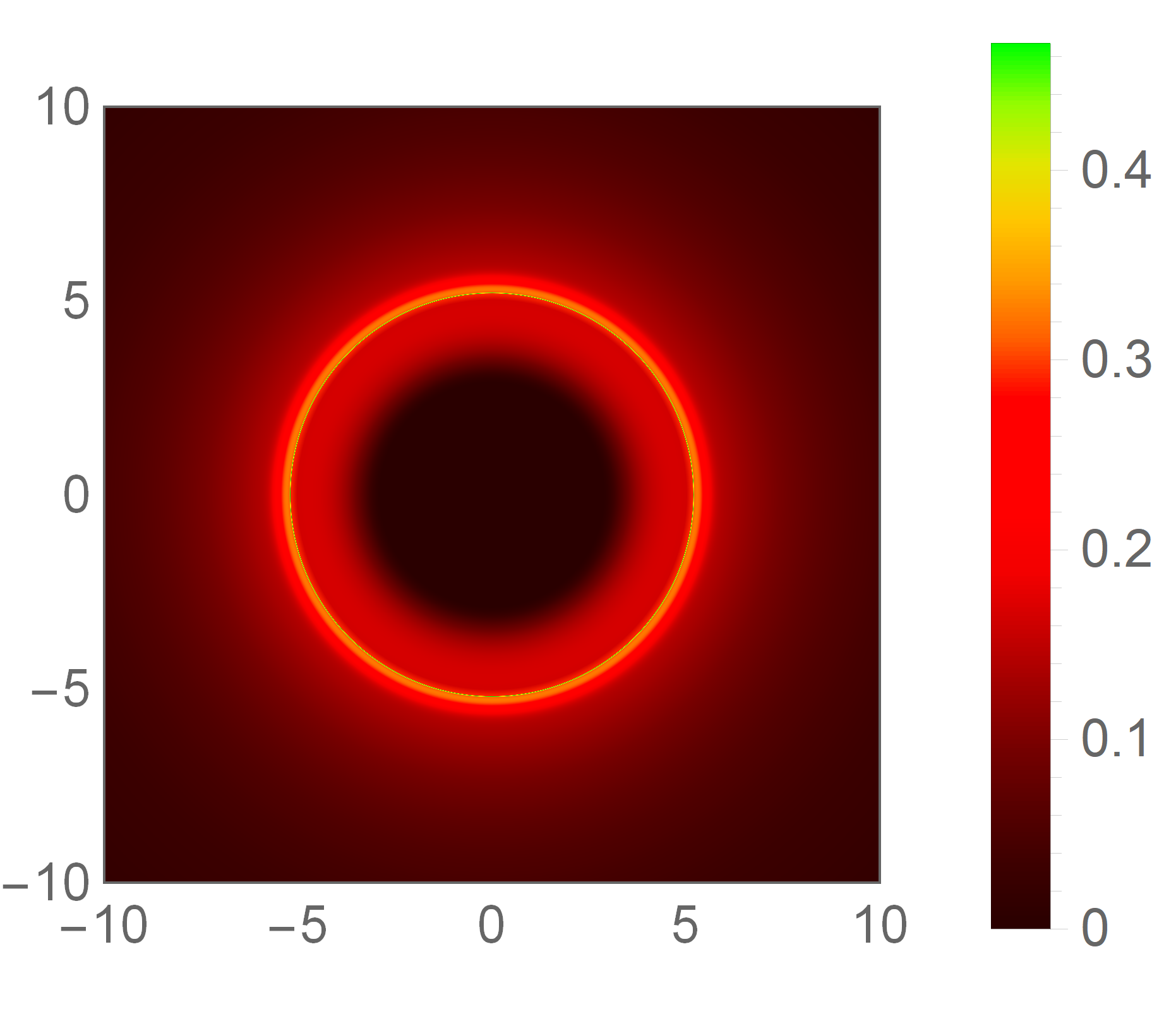}
\includegraphics[width=4.4cm,height=4cm]{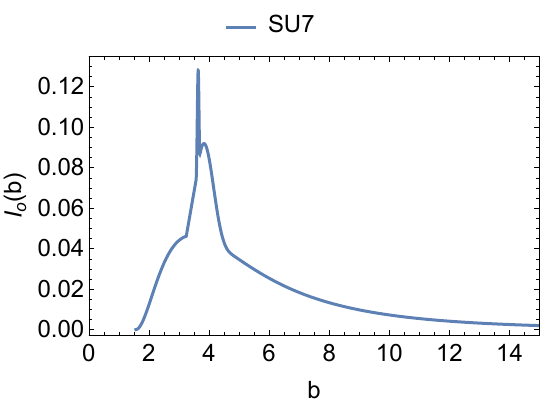}
\includegraphics[width=4.4cm,height=4cm]{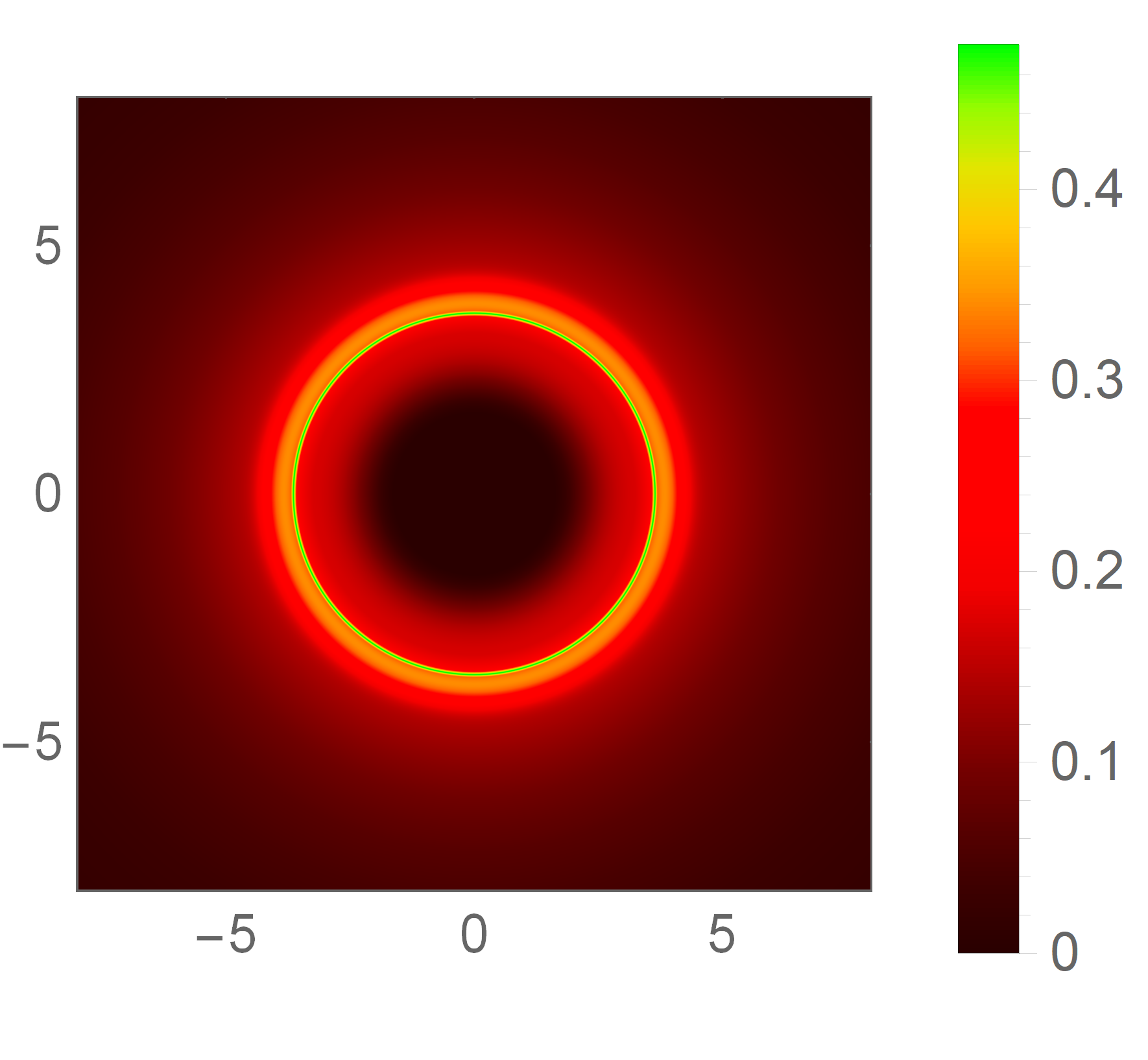} \\
\includegraphics[width=4.4cm,height=4cm]{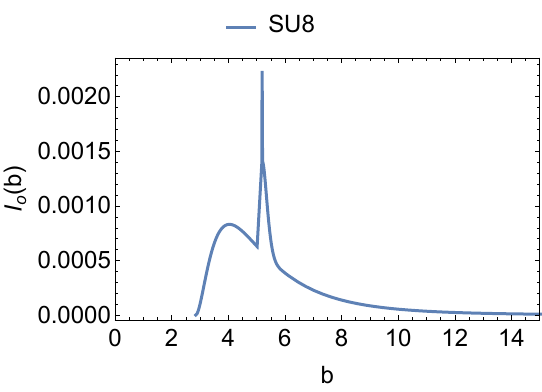}
\includegraphics[width=4.4cm,height=4cm]{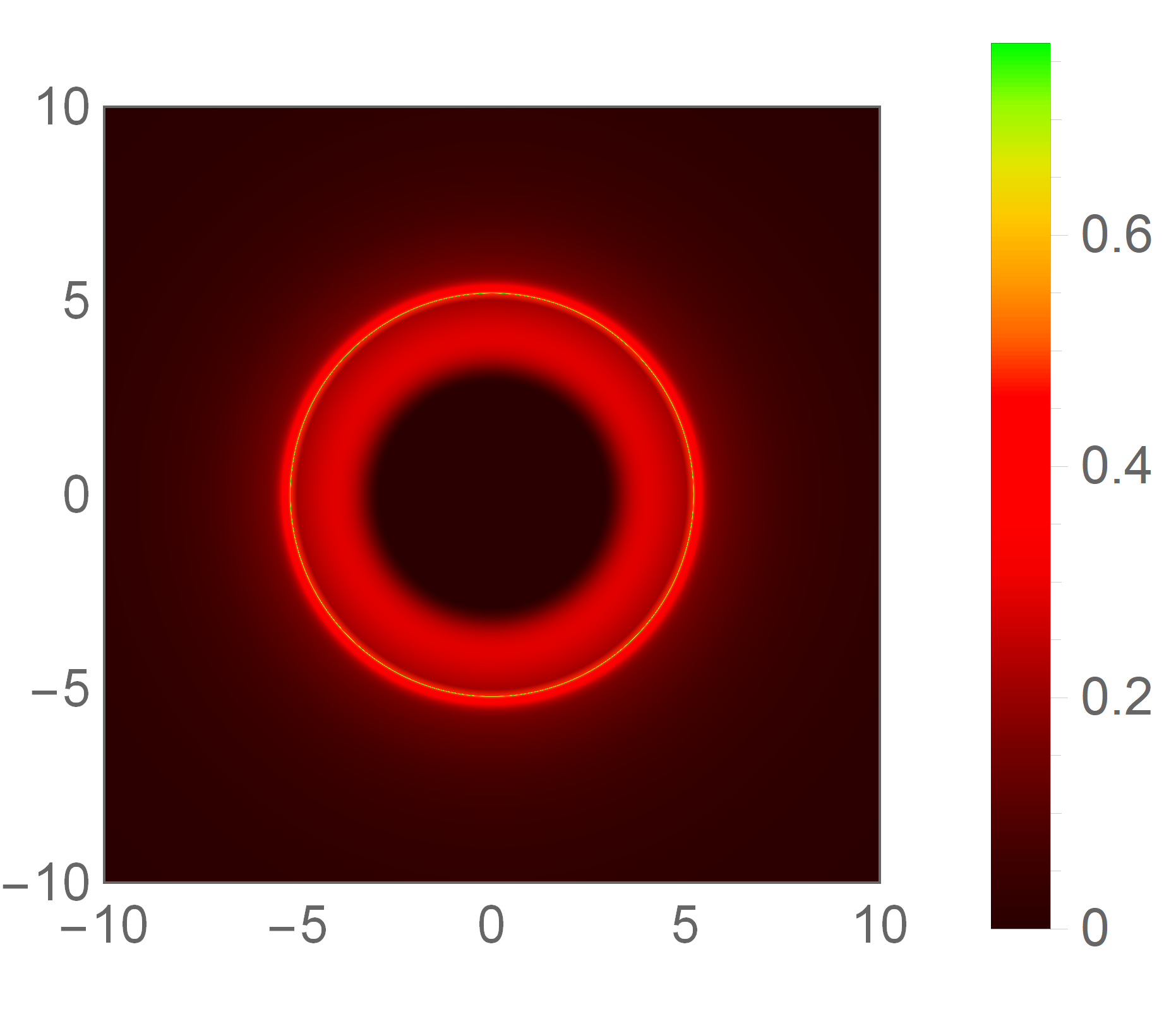}
\includegraphics[width=4.4cm,height=4cm]{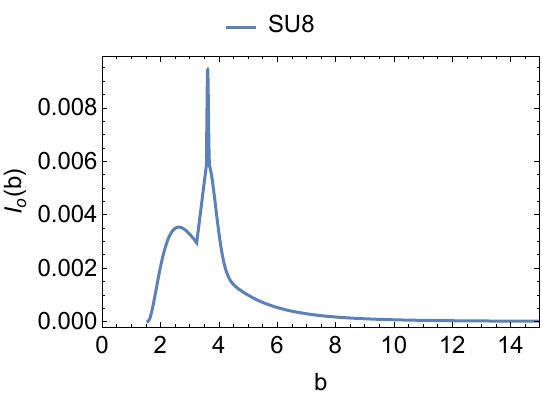}
\includegraphics[width=4.4cm,height=4cm]{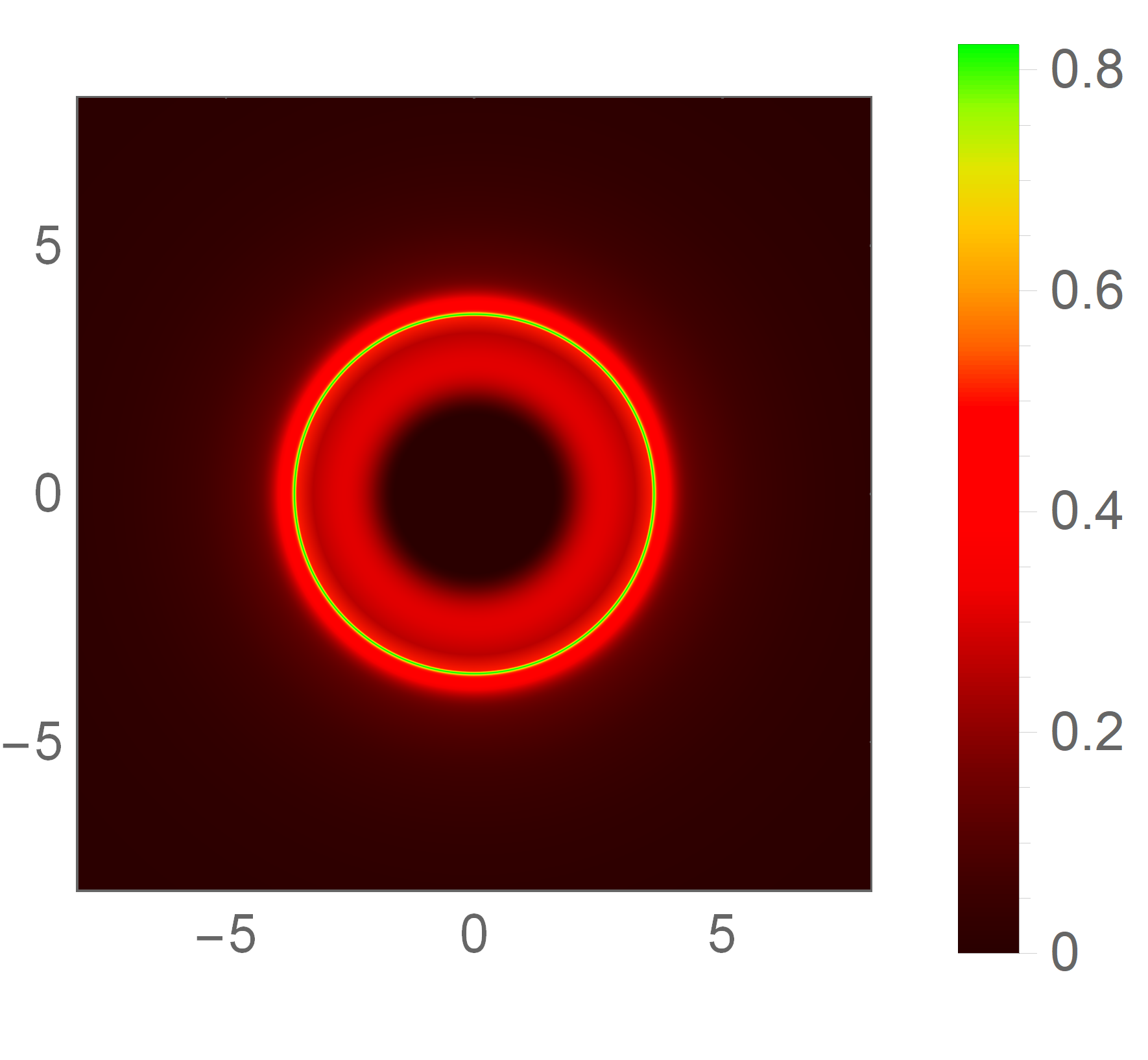} \\
\includegraphics[width=4.4cm,height=4cm]{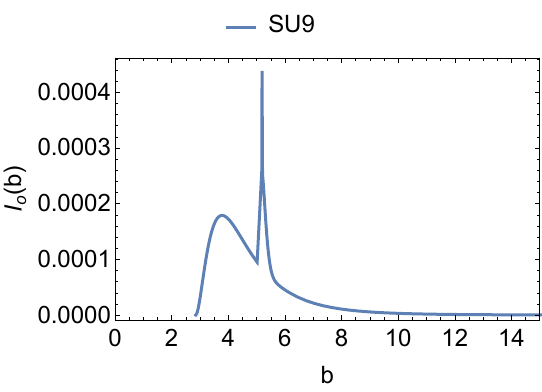}
\includegraphics[width=4.4cm,height=4cm]{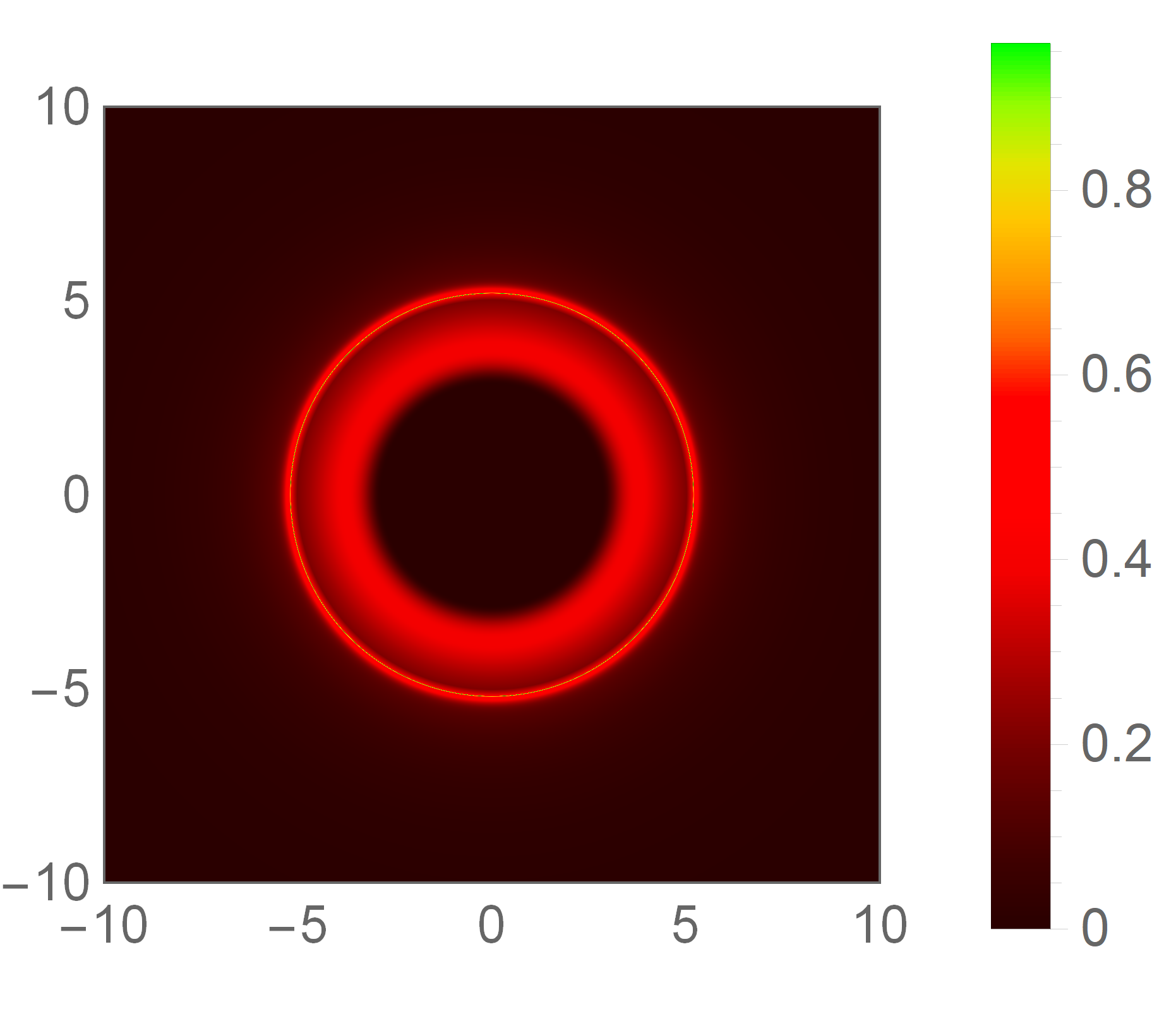}
\includegraphics[width=4.4cm,height=4cm]{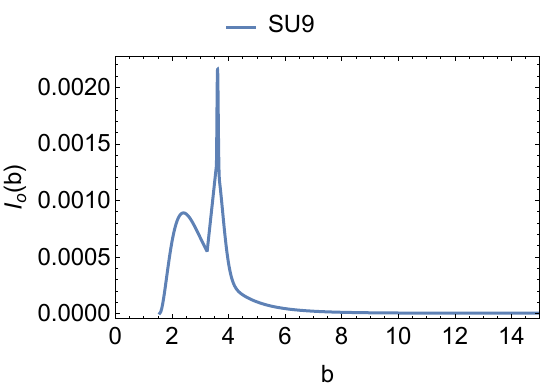}
\includegraphics[width=4.4cm,height=4cm]{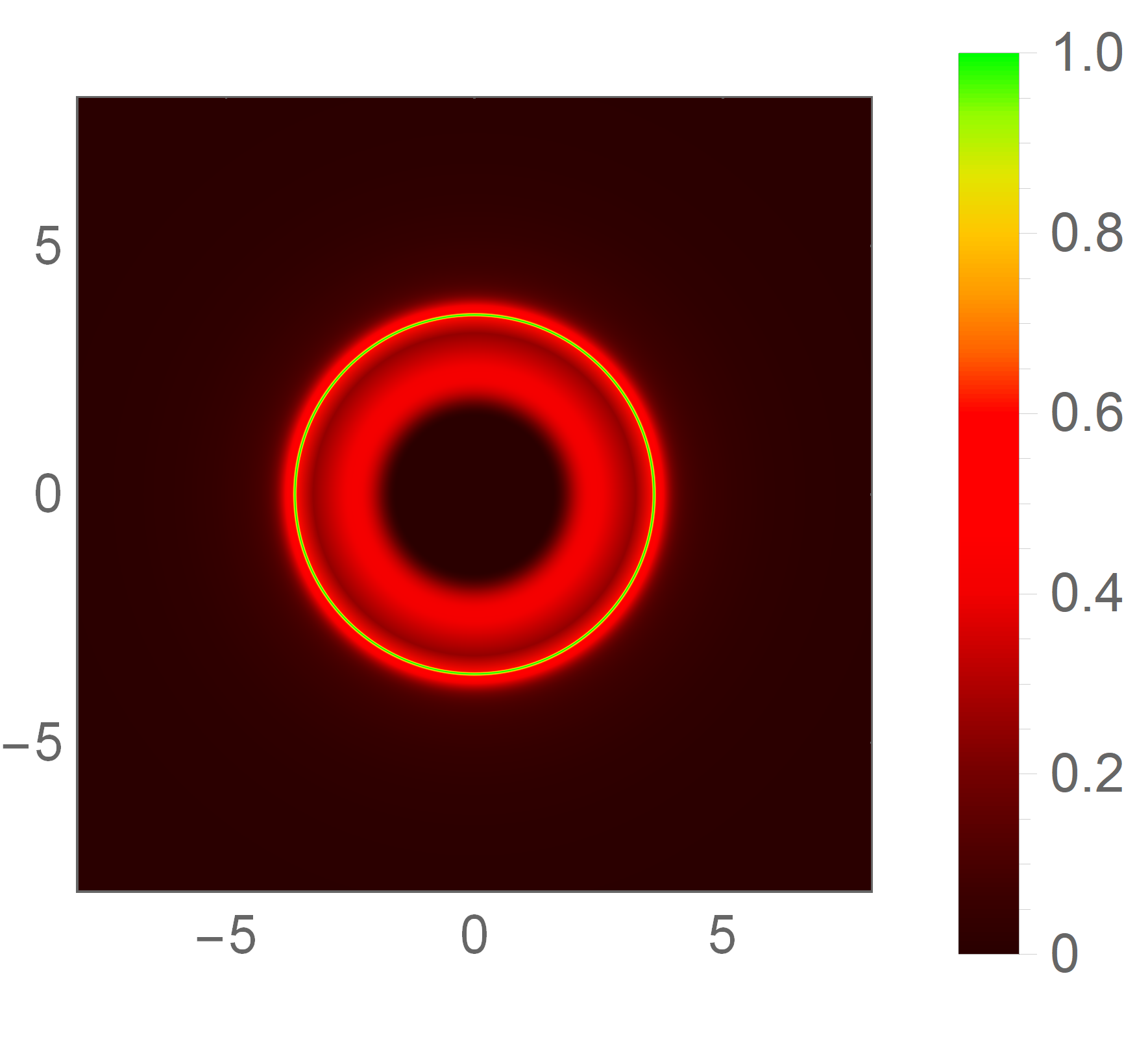} \\
\includegraphics[width=4.4cm,height=4cm]{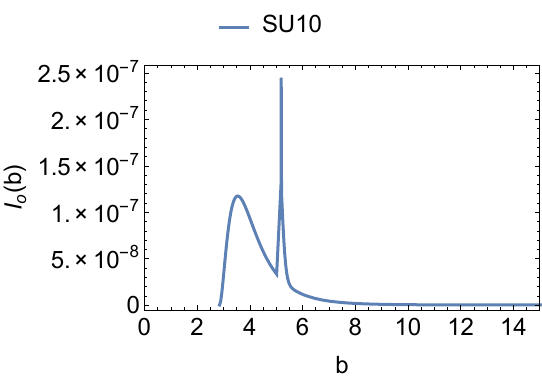}
\includegraphics[width=4.4cm,height=4cm]{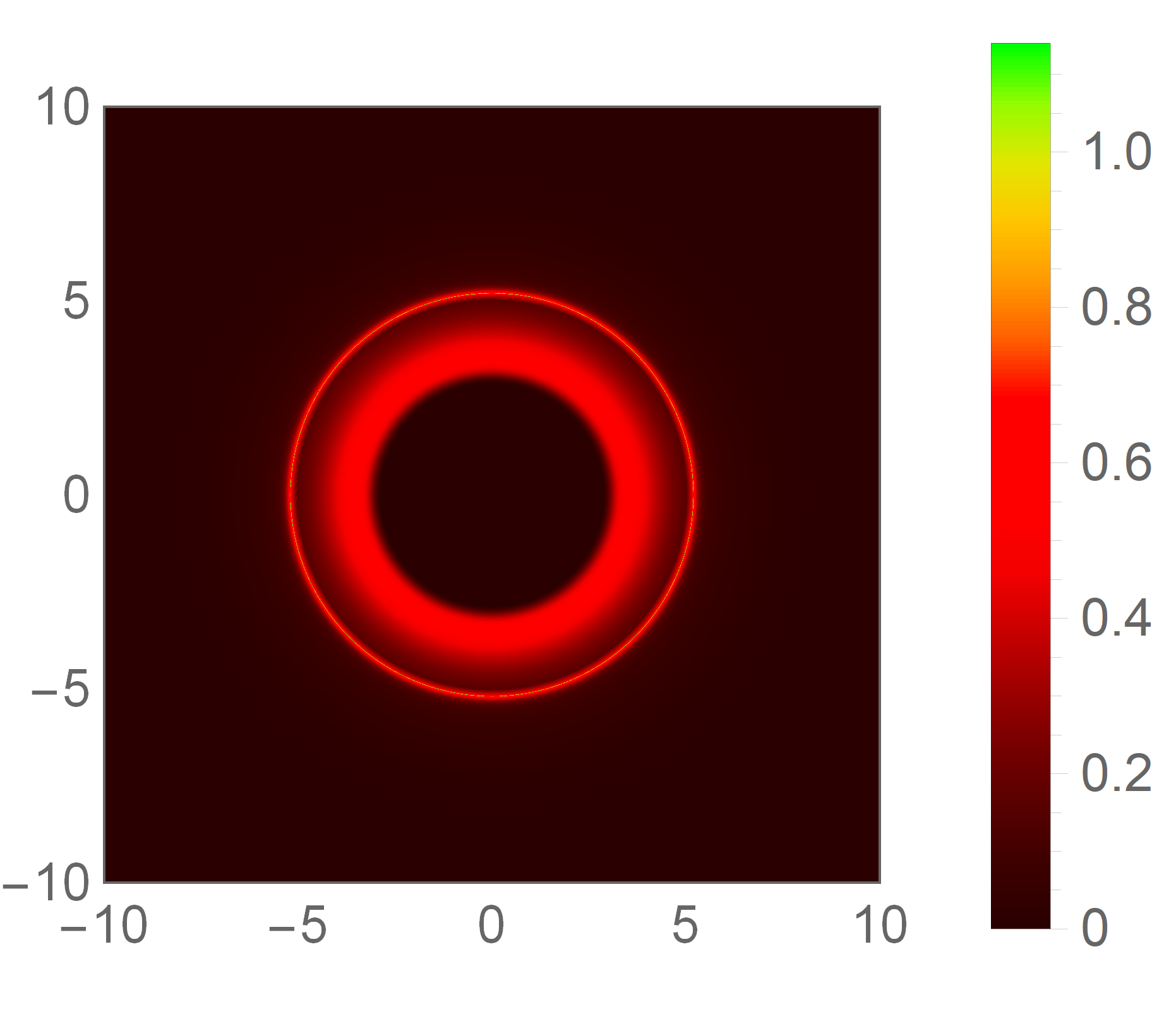}
\includegraphics[width=4.4cm,height=4cm]{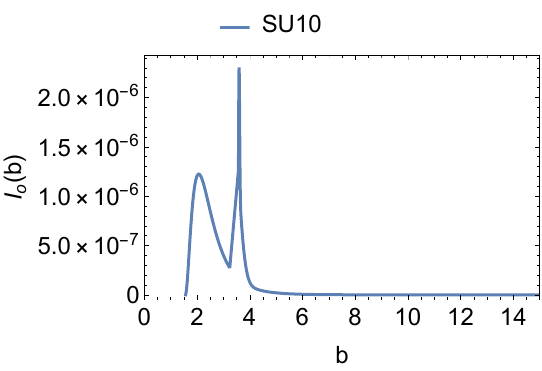}
\includegraphics[width=4.4cm,height=4cm]{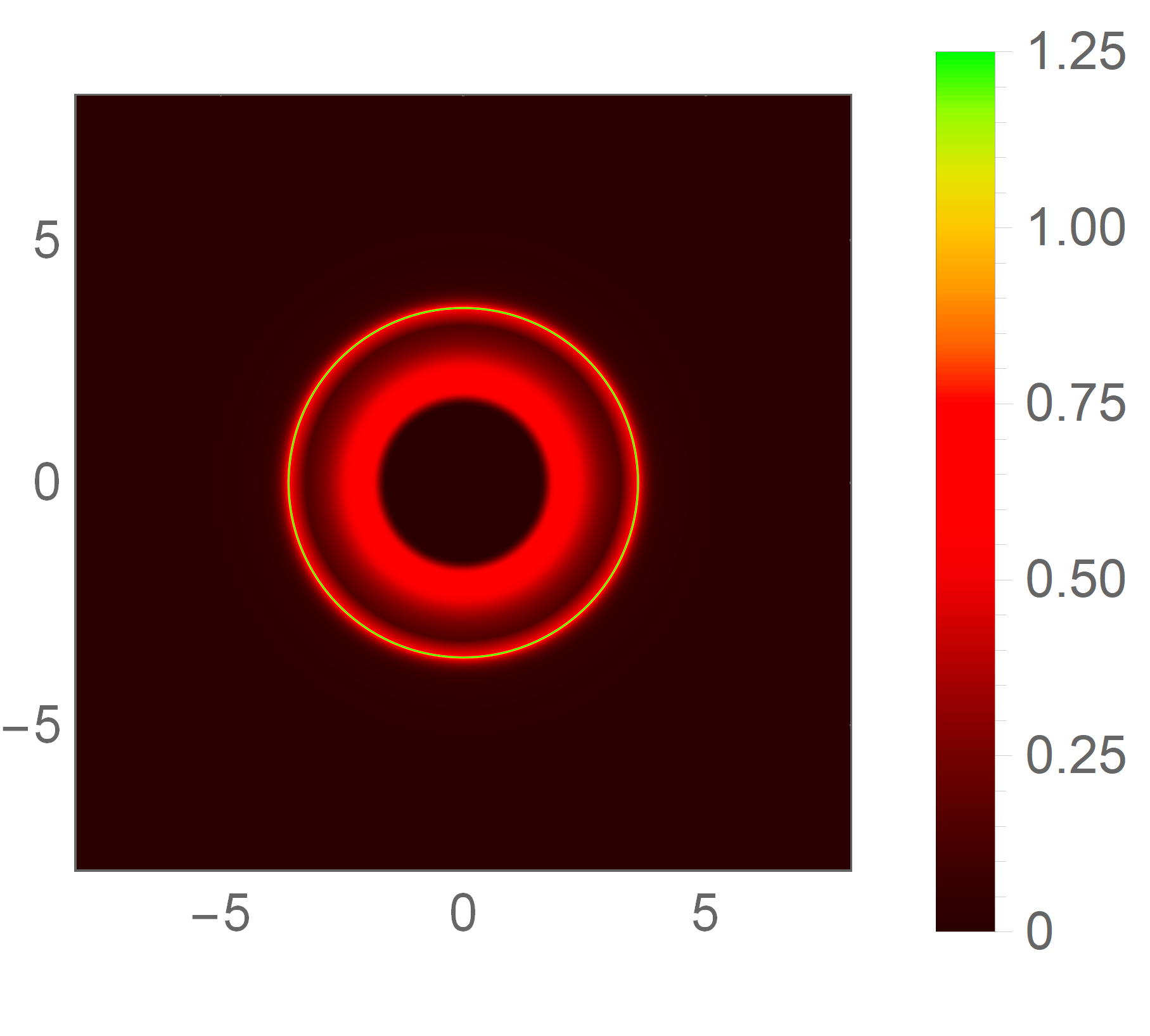}
\caption{The observed intensity $I_{o}(b)$ and the optical appearance (using $b \in [-8,8]$) of a Schwarzschild black hole (left two figures) and a dSBH solution (right two figures) for the (from top to bottom) SU6, SU7, SU8, SU9, and SU10 emission models of Table \ref{Table:I}, using a fudge factor $\xi_0=\xi_1=\xi_2=1$ and $\xi_n=0$ for $n>2$.}
\label{fig:images2}
\end{figure*}

In Table \ref{Table:II}, we report the numerical results of our simulations for the Schwarzschild and dSBH configurations. Our main  interest here is to check the reliability of the Lyapunov index $\gamma_L$ of Eq.(\ref{eq:Lya}) as a marker of the actual exponential decay of the luminosity of successive photon rings. To this end, we compute the extinction rate of the $n=1$ ring as compared to the $n=2$, that is, the luminosity ratio $I_1/I_2$ for each SU model, and compare it to the theoretical rate $e^{\gamma_L}$\footnote{For the latter we do not use the $n \to \infty$ limit where (for the Schwarzschild black hole) $\gamma_L =\pi$, but approximate it by the $n=2$ value (for Schwarzschild $\gamma_L \approx 3.150$). This difference ($\sim 0.3  \%$) is negligible enough for our purposes here .}. From this Table, we observe that in both the Schwarzschild and dSBH geometries five SU models produce higher extinction rates than its Lyapunov-based rate, one is left unchanged, while the other four predict a lower rate. Such deviations from the theoretical prediction can be quite important: in the SU1 model it can increase up to a $\sim 24\%$ for Schwarzschild, while in the SU3/SU6 it can decrease by a factor $\sim 10 \%$ for Schwarzschild and $\sim 15 \%$ for dSBH. This behaviour is strongly governed by the disk's parameter $\gamma$: negative (positive) values of $\gamma$ yield larger (smaller) extinction rates; in this sense, the SU5 and SU8 models, having $\gamma=0$, produce extinction rates that track very closely the Lyapunov prediction. This is in agreement with the results of GRMHD simulations, which associate the size of $\gamma$ with the thickness of the rings and, in turn, with their relative luminosity, as we have just seen here.

In Figs. \ref{fig:images1} and \ref{fig:images2} we depict the observed intensity $I_{o}(r)$ and the optical appearance for a range of values for the impact parameter  $b \in [-8,8]$ for the Schwarzschild (left two plots) and dSBH (right two plots) solutions, and for the SU1/SU2/SU3/SU4/SU5 and SU6/SU7/SU8/SU9/SU10 models, respectively, and we recall that such models are organized in decreasing values of their effective region of emission. There are several features of interest in these images. We find universally (\textit{i.e.} for every SU model, and in both the Schwarzschild and dSBH configurations) the expected presence of both a central brightness depression and a bright photon ring surrounding it, which can be further decomposed into up to two additional rings (the photon rings) depending mostly on the choice of SU model but also with some differences between the Schwarzschild/dSBH solutions.

Regarding the bright rings, their distribution is governed by both the location of the effective source of emission and the strength of the decay with distance of the emission profile. We can see this effect in the SU1/SU2 models, where the location of the maximum of emission is very similar but the decay is weak (strong) in the SU1 (SU2) model: as a consequence in the former (latter) the glow of the direct emission extends to much farther (closer) regions of the impact parameter space. The photon rings ($n=1,n=2$) will be clearly visible in the Schwarzschild-SU2 model, but they appear overlapped with the direct emission in the Schwarzschild-SU1 and dSBH-SU1/SU2 models, which means we are able to distinguish between both background geometries at fixed SU emission model. Both SU3 and SU6 produce an unusual signature for both Schwarzschild/dSBH models with the $n=1$ photon ring located above the direct emission but separated from it, so in the optical appearance one neatly sees the photon ring  encircling the direct emission; this effect is more exaggerated in the SU3 model given the fact that the decay of the profile with distance is stronger than in SU6. In every other SU model the photon rings are overlapped with the direct emission. In particular, for the SU4 it nears the outer edge of the brightness depression while the direct emission extends to farther distances given the weak decay of the intensity profile with the distance (as in SU1). Finally, SU5, SU7, SU8, SU9, and SU10 produce quite similar images: the direct emission dominates the image from the central brightness depression outwards, and superimposed on it lies the boost of luminosity of the two photon rings, whose exact locations within the direct emission depend on the strength of the decay of the intensity profile which defines the effective domain of the direct emission: weak (strong) decays produce a photon ring placed far (near) the outer edge of the image.

\begin{figure*}[t!]
\includegraphics[width=5.4cm,height=4.6cm]{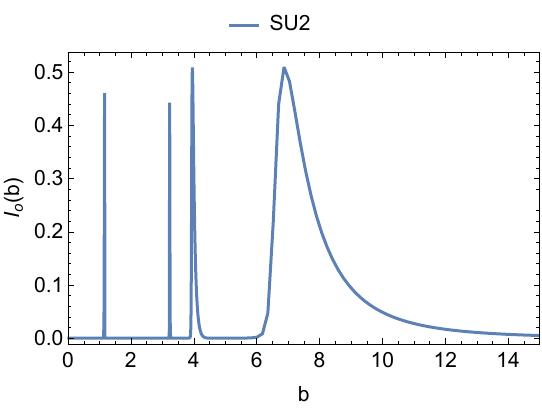}
\includegraphics[width=5.4cm,height=4.6cm]{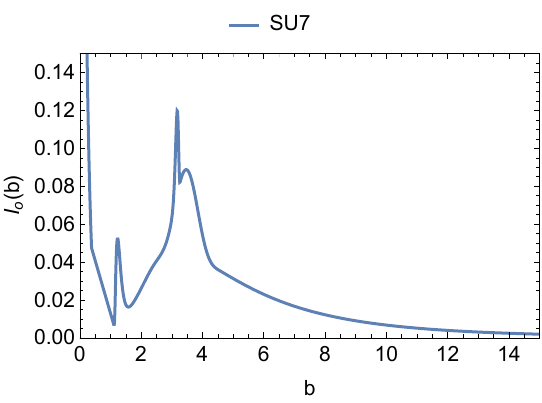}
\includegraphics[width=5.4cm,height=4.6cm]{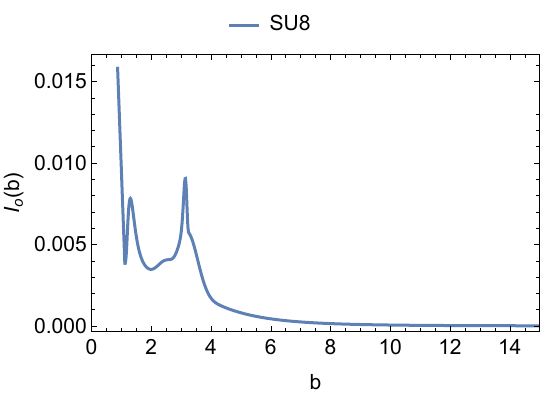} \\
\includegraphics[width=5.4cm,height=4.6cm]{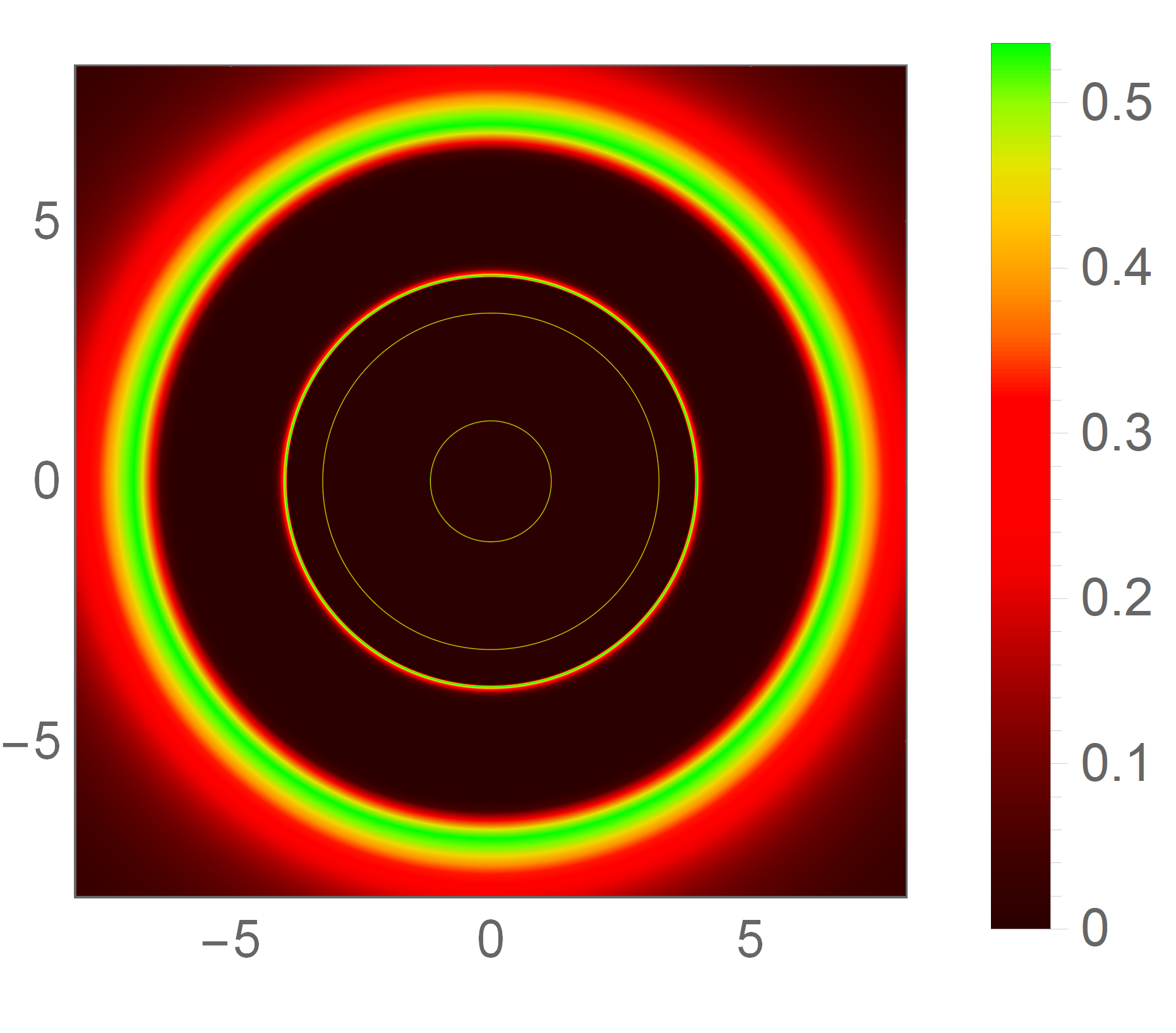}
\includegraphics[width=5.4cm,height=4.6cm]{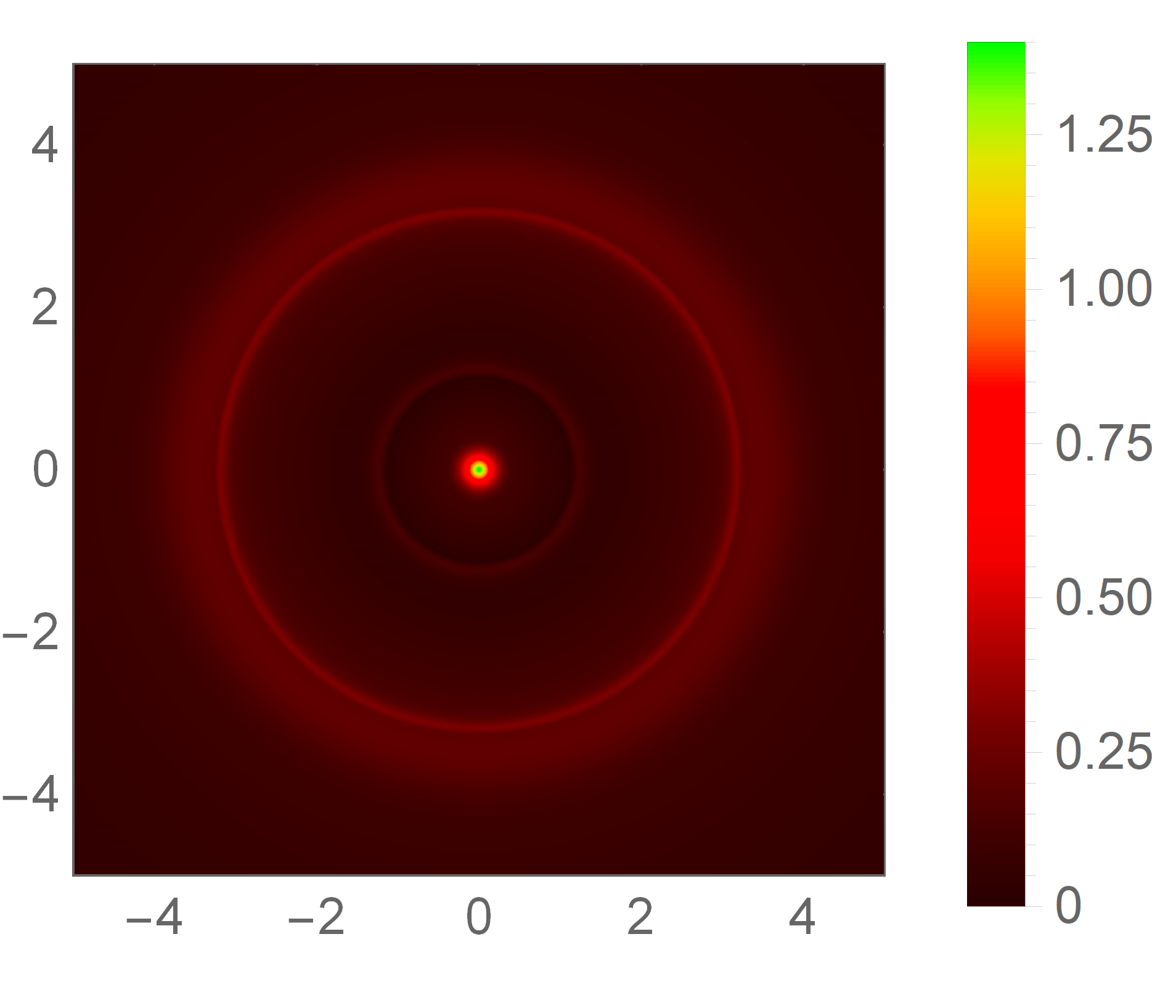}
\includegraphics[width=5.4cm,height=4.6cm]{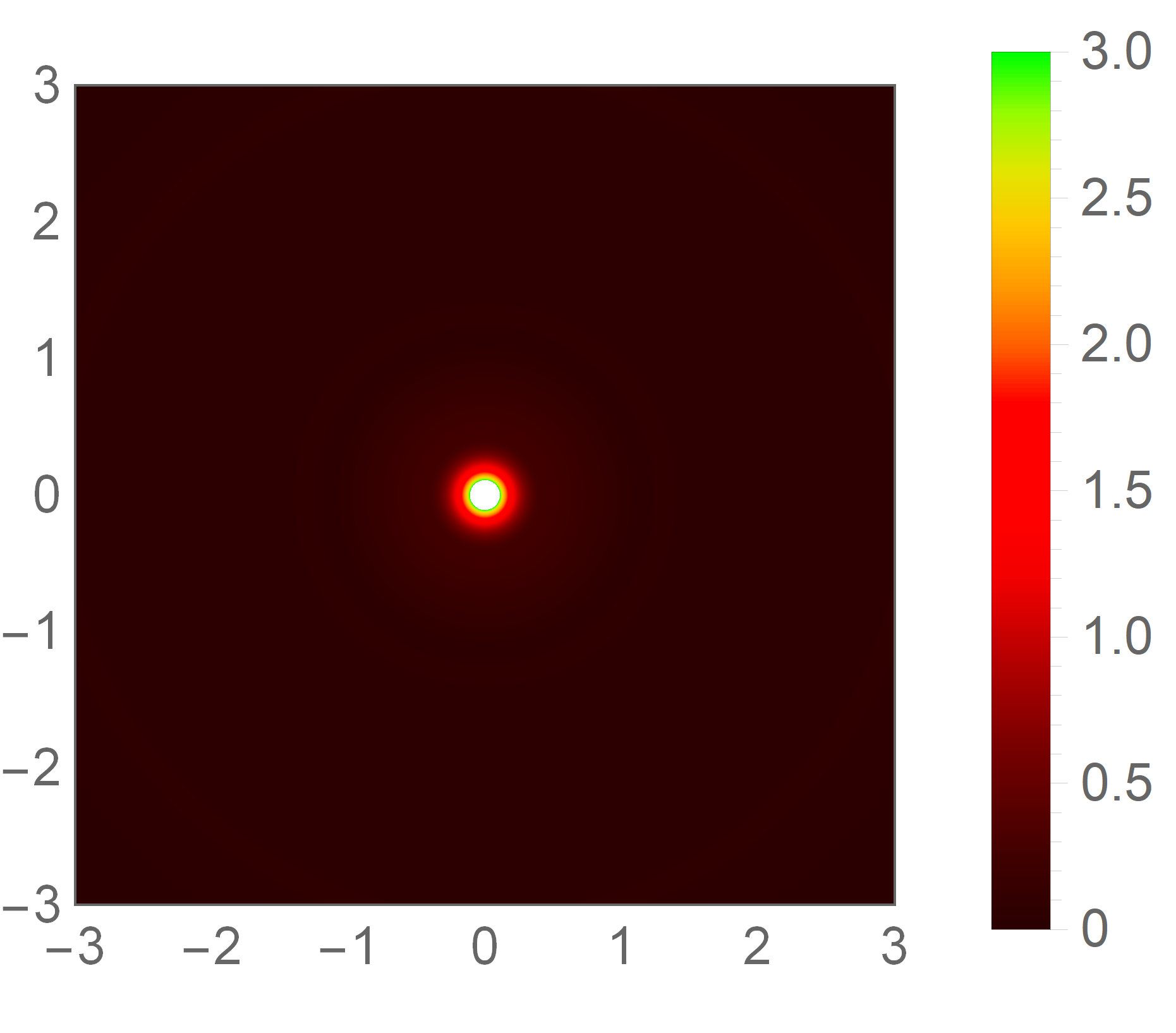}
\caption{The observed intensity $I_{o}(r)$ (top) and the optical appearance (bottom) of a naked regular dS core for the SU2 (left), SU7 (middle), and SU8 (right) emission models of Table \ref{Table:I}, using a fudge factor  $\xi_0=\xi_1=\xi_2=1$ and $\xi_n=0$ for $n>2$.}
\label{fig:images3}
\end{figure*}

Differences between the dSBH and Schwarzschild solutions (at fixed SU model) are both obvious at naked eye but also subtle. The first significant difference is a strong reduction of the central brightness depression size in the dSBH case as compared to the Schwarzschild one. This was expected on the grounds of a much smaller horizon radius, critical impact parameter and photon sphere radius of the dSBH, since these aspects strongly influence the actual location of the outer edge of the brightness depression, and which in the case where the emission goes all the way down to the event horizon is bounded by above by the so-called {\it inner shadow},  and which, as already discussed, depends only on the background geometry \cite{Chael:2021rjo}. The second difference is with the photon rings themselves: we already computed in Table \ref{Table:II} a significant reduction in the extinction rate $I_1/I_2$ between both geometries (up to a factor 2), which has a clear reflection of brighter rings as compared to the surrounding image. Besides these aspects, the arrangement of the locations and overlapping of the photon rings with the direct emission in the dSBH has some tiny differences with respect to the Schwarzschild case but mostly in those SU models with emission outside the event horizon; those disks whose emission decreases from the horizon outwards tend to wash out these differences between background geometries. Let us recall that these images are obtained in a face-on orientation, $i=0$: inclined images would make the lower part of the photon ring to tilt upon the central brightness depression introducing an asymmetry in the images (though only noticeable at large enough inclinations). However, this effect would be similar for both Schwarzschild and dSBH and would help little to distinguish them beyond what we discussed here; an illustration of this effect can be found in e.g. \cite{daSilva:2023jxa}.

\subsection{Regular naked dS core}

For the sake of completeness of our analysis, let us briefly discuss an example of a regular naked dS core, for which we choose the value $l=0.30$, which brings it very close to becoming a (extreme) black hole. In this case, the lack of a horizon allows light trajectories to interact with the inner part of the effective potential, which has a divergent behaviour at the center. This has a dramatic effect in the images of the corresponding objects. To illustrate this, we depict in Fig. \ref{fig:images3} the observed intensity and the optical appearances for three choices of the SU models (SU2, SU7, and SU8), corresponding to the GLM3/GLM1/GLM2 used in previous papers by some of us. There one neatly sees the main difference brought by these NdS objects: the appearance of new peaks of intensity, which manifest themselves as new photon rings in the optical appearances. The reason for this can be seen in the transfer function of Fig. \ref{fig:transfunc} (right plot), where photons interacting with the internal part of the potential will be repelled by it and have the chance to circulate one more time around the photon sphere, producing the new sets of photon rings in the observer's screen. Furthermore, such new contributions break the expected exponentially decay of successive photon rings, which makes rings with $n >2$ to produce non-negligible contributions to the luminosity of the images, and yield a multi-ring structure previously studied in the literature \cite{Peng:2021osd,Guerrero:2022msp}, though for simplicity we just kept up to the $n=2$ ring to illustrate the main aspects of the optical appearances of the NdS solutions. Indeed, the latter resemble the canonical appearance of a black hole for the SU2 model, with the new photon rings living inside the region previously occupied by the central brightness depression only, yielding a clear signature to distinguish both objects. But in the SU7 and SU8 models the resulting objects yield optical appearances that have nothing to do with the canonical black hole ones and, as such, these kinds of objects can hardly pass as black hole mimickers from what we know about their cast images.

\section{Conclusion}		 \label{Sec:IV}

In this work we have considered a black hole geometry replacing the central region of the Schwarzschild black hole by a de Sitter core characterized by a single parameter $l$, this way removing the curvature singularity lurking there, and studied its optical appearance when surrounded by an optically and geometrically thin accretion disk. This de Sitter core is actually the strongest modification of the Schwarzschild solution of a certain class of such cores which include the well known Bardeen and Hayward solutions as particular members.

We set a value of $l=0.25M$ that is allowed by the recent analysis of the orbital motion of the S2 star around Sgr A*, that estimated $l \lesssim 0.47M$ at the 95\% of the confidence level, and corresponds to the lower end of the sub-family of these black holes which are thermodynamically stable. Regarding the accretion disk, we use a simplified setting for a monochromatic intensity profile belonging to the class of models known as Standard Unbound Johnson's distribution, a sub-class of which has been shown in the literature to be capable to reproduce the results of some scenarios for the accretion flow in GRMHD simulations. In particular, we employed ten choices for the parameters of such profiles, representatives of different emission profiles and producing different signatures of the corresponding optical appearances.

We run our simulations in this setting, which for the background geometry means significantly lower horizon radius, critical impact parameter, and photon sphere radius. This is translated, within the optical appearances of these objects, into a strong suppression of the luminosity extinction rate (up to a factor two) between the $n=1$ and $n=2$ photon rings, which therefore appear in the images as significantly more luminous within the direct emission of the disk as compared to their Schwarzschild counterparts. This happens for all the emission models, yet visible differences between dSBH/Schwarschild configurations are more acute in those SU model whose peak of emission is located away from the event horizon; those in which the profile decreases from the event horizon outwards tend to wash out such differences and produce more similar images.

Furthermore, the dS black hole also yields a strongly reduced size of the outer edge of the central brightness depression, neatly distinguishing it from the Schwarzschild black hole. Under the recent results from the calibrated measurements of the ``shadow's size" of Sgr A$^*$ by the EHT Collaboration reported in \cite{EventHorizonTelescope:2022xqj}, where it is claimed that the size of the bright ring correlates, under certain  assumptions and after proper calibration (and some debate) with such an edge, one expects the latter to be constrained (at $2\sigma$) within the range $4.21 \lesssim r_{sh}/M \lesssim 5.66$, while in our case this is $r_{sh}/M \approx 3.614$. This strategy was actually used in \cite{Vagnozzi:2022moj} to constrain a large number of alternative spherically symmetric geometries (further analysis of the implications of such a measurement has been recently discussed in \cite{Broderick:2023jfl}). Given the strong reduction of the size of the central brightness depression in our case, this makes these objects hardly compatible with the images of Sgr A$^*$, pinpointing that strong deviations from the GR solutions may have a hard time in reconciling themselves with present and future observations. Obviously, such a criticism applies even more strongly to the naked dS configurations since they also show a multi-ring structure that does not resemble current  observations, except for very particular emission model scenarios.

To conclude, our analysis with a pool of well motivated emission models shows that regular de Sitter-core black holes can be actually tested with current and feature ``shadow" observations  since they introduce large enough differences in their optical appearances as compared to canonical black hole solutions. Even though the background geometry employed here seems to be too strong a modification of the Schwarzschild one to be compatible with current images, it paves the way for other dS core black holes to be tested with this tool. Of course, in order to produce competitive images, the simplified treatment employed here needs to be upgraded towards more realistic settings with rotation, thick disks, absorption, inclination, and so on.

\section*{Acknowledgments}

IDM and RDM acknowledge  support from the  grant PID2021-122938NB-I00, and DRG acknowledges support from the grant PID2022-138607NB-I00, both funded by MCIN/AEI/10.13039/501100011033 and by ``ERDF A way of making Europe''.

\appendix
\section{Null geodesic behaviour}  \label{S:app}

In this Appendix we provide a short summary of the main ingredients (since they have been derived many times in the literature of the subject, see e.g. \cite{Perlick:2021aok} for a quite general analysis) of the equations of null geodesic motion together with their main features for the generation of images. We start from the Lagrangian density of a particle, which is given by
\begin{equation}
\mathcal{L}=\frac{1}{2} g_{\mu\nu}\dot{x}^{\mu}\dot{x}^{\nu} \ ,
\end{equation}
where a dot represents a derivative with respect to the affine parameter. For a photon with wave function $k^{\mu}=\dot{x}^u$ this Lagrangian equals zero, a reflection of the fact that photons travel along null geodesics of the background geometry, i.e., $g_{\mu\nu} k^{\mu}k^{\nu}=0$. The spherically symmetric character of the geometry means that we can fix $\theta=\pi/2$ without any loss of generality, while the existence of two Killing vectors associated to translational and rotational invariances of the system  implies two conserved quantities, $E=-A\dot{t}$ and $L=r^2 \dot{\phi}$, naturally interpreted as the energy and angular momentum per unit mass, respectively. This allows to write the geodesic equation under the form
\begin{equation} \label{eq:geonull}
\dot{r}^2=\frac{1}{b^2} -V(r) \ ,
\end{equation}
where $b=L/E$ is the impact parameter, while the effective potential reads as
\begin{equation}
V(r)=\frac{A(r)}{r^2} \ .
\end{equation}
Eq.(\ref{eq:geonull}) has the typical shape of a particle moving in a one-dimensional effective potential. From here one can define the photon sphere as given by the simultaneous fulfilment of the equations
\begin{eqnarray}
&&b_c^2 =\frac{1}{V(r_m)}  \\
&&V'(r)\vert_{r=r_m}=0 \\
&&V''(r)\vert_{r=r_m}<0 \ .
\end{eqnarray}
The first condition states that a turning point has been reached, the second its critical (extremum) character, and the third that it corresponds to a maximum (otherwise, it is dubbed as an anti-photon sphere, which are typically unstable \cite{Cardoso:2014sna}). Provided that all these conditions are met then the photon sphere radius $r_m$ (which is projected in the observer's screen as a circular critical curve) yields the locus of bound unstable geodesics, with $b_c$ corresponding to the {\it critical impact parameter} a light ray issued from the observer's screen would need in order to asymptote to $r=r_m$. This way, $b_c$ separates those trajectories with $b>b_c$ that find a turning point at some $r_t>r_m$, from those with $b<b_c$ that overcome the effective potential and end up swallowed by the event horizon of the black hole. In the backwards ray-tracing procedure, it splits the observer's plane image into the bright ($b>b_c$) and dark ($b<b_c$) regions, with those nearing $b \gtrsim b_c$ being associated to the presence of photon rings, namely, light trajectories that have circled the black hole $n$-half times. In order to classify the latter and to find the optical appearance of the object one is interested in obtaining the deflection angle; this is done using again the conserved quantities of the system to rewrite (\ref{eq:geonull}) as a variation of the azimuthal angle $\phi$ with respect to $r$, which results in Eq.(\ref{eq:dphidr}). Nonetheless, in thin disk scenarios where emission from inside the photon sphere (up to the event horizon) can occur, this simple picture must be modified in order to allow light rays to get out from the accretion disk towards reaching the asymptotic observer; this is described in the main text.

\end{document}